\documentclass[final,5p,times,twocolumn]{elsarticle}
\usepackage{amsmath,amssymb}
\usepackage{color}
\usepackage{subfig}
\usepackage{graphicx}
\usepackage{tabularx}
\usepackage{lineno}
\usepackage[version=4]{mhchem}
\modulolinenumbers[5]

\usepackage[euler]{textgreek}
\usepackage{siunitx}
\DeclareSIUnit{\wtpercent}{wt.\%} 
\DeclareSIUnit{\atpercent}{at.\%} 
\usepackage{miller}

\journal{Corros. Sci.}

\begin{document}

\begin{frontmatter}

\title{AgCl-induced hot salt stress corrosion cracking in a titanium alloy}

\author[IC]{Yitong Shi}
\author[IC]{Sudha Joseph}
\author[RR2]{Edward A. Saunders}
\author[RR1]{Rebecca S. Sandala}
\author[RR1a]{Adrian Walker}
\author[IC]{Trevor C. Lindley}
\author[IC]{David Dye}

\address[IC]{Department of Materials, Royal School of Mines, Imperial College London, Prince Consort Road, London, SW7 2BP, UK}
\address[RR2]{Rolls-Royce plc., Materials - Failure Investigation, Bristol BS34 7QE, UK}
\address[RR1]{Rolls-Royce plc., Elton Road, Derby, DE24 8BJ, UK}
\address[RR1a]{Retired; formerly with Rolls-Royce plc., Elton Road, Derby, DE24 8BJ, UK}

\begin{abstract}
The mechanism of AgCl-induced stress corrosion cracking of Ti-6246 was examined at \SI{500}{\mega\pascal} and \SI{380}{\celsius} for \SI{24}{\hour} exposures. SEM and STEM-EDX examination of a FIB-sectioned blister and crack showed that metallic Ag was formed and migrated along the crack. TEM analysis also revealed the presence of \ce{SnO2} and \ce{Al2O3} corrosion products mixed into \ce{TiO2}. The fracture surface has a \textcolor{black}{transgranular nature with a brittle appearance in the primary} $\alpha$ phase. Long, straight and non-interacting  dislocations were observed in a brittle appearance fractured primary $\alpha$ grain, with basal and pyramidal traces. This is consistent with a dislocation emission view of the the cracking mechanism.
\end{abstract}

\begin{keyword}
Titanium alloys \sep Stress corrosion cracking (SCC) \sep STEM-EDX\sep Hydrogen embrittlement \sep Solid metal embrittlement
\end{keyword}

\end{frontmatter}

\section{Introduction}
\label{Intro}


 Titanium alloys generally exhibit good corrosion resistance owing to the formation of a well-adhered and protective nanometric TiO$_2$ layer \cite{lutjering2007titanium}. Combined with the benefits of low density and good intermediate-temperature mechanical properties, Ti-based alloys are found to be outstanding structural materials for weight-critical fatigue limited aerospace applications at elevated temperatures (\SI{300}{\celsius}-\SI{600}{\celsius}) \cite{boyer1996overview}. Approximately one third of the structural weight of modern turbofan engines are composed of titanium components, mostly in the fan and compressor sections, as both blades and discs \cite{peters2003titanium}. Ti-6Al-2Sn-4Zr-6Mo (wt.\%, Ti-6246), is a relatively heavily $\beta$-stabilised $\alpha$ + $\beta$ alloy used in high temperature compressor discs/rotors due to its good elevated temperature strength \cite{boyer1996overview}.


 However, since the 1960s there has been a recurring concern around hot salt stress corrosion cracking (HSSCC) susceptibility in Ti alloys when exposed to halides, particularly chlorides \cite{rideout1966basic,turley1966elevated,simenz1966environmental,beck1968stress,blackburn1969metallurgical}. Bauer first reported the failure of a Ti-6Al-4V turbine blade due to HSSCC attack by NaCl residues from fingerprints during laboratory creep testing at  \SI{220}{\celsius}~\cite{Petersen1971}. Subsequent investigations found that other chloride-containing metal salts, such as MgCl$_2$, CaCl$_2$ and KCl, found in seawater could cause a similar effect \cite{rideout1966basic,turley1966elevated}.   Molten salts can also cause liquid metal embrittlement (LME) and have more deleterious effects on mechanical properties \cite{russellh.jones2017}.  In addition, the susceptibility to HSSCC in titanium alloys can be influenced by test conditions, alloy composition and alloy heat treatment condition \cite{simenz1966environmental,beck1968stress}. The threshold stress intensity for SCC, K$_\mathrm{lSCC}$,  is well below both the conventional fatigue cracking threshold stress intensity and fracture toughness. K$_\mathrm{lSCC}$ is dramatically reduced by increasing the exposure time and temperature according to a Larson-Miller type relationship \cite{turley1966elevated}.
 
 A number of reaction models or mechanisms have been proposed and it is widely agreed that the presence of moisture and/or oxygen is critical for cracking 
 \cite{rideout1966basic,Petersen1971,garfinkle1973electrochemical}. Initially the protective oxide layer can be ruptured mechanically or consumed through its reactions with salts in the presence of moisture and/or oxygen, subsequently forming HCl or Cl$_2$. Then the underlying alloy can be attacked by HCl or Cl$_2$, generating  titanium chlorides and atomic hydrogen at elevated temperatures, $>$ 
 \SI{300}{\celsius}. Other alloy chlorides have also been 
 observed, such as Al, Sn and Zr. Pyrohydrolysis reactions of such chlorides can provide further HCl which can then 
 continue to attack the base alloy. Titanium hydrides have been observed by X-ray diffraction (XRD) and transmission electron microscopy (TEM) \cite{joseph2018mechanisms}.
Consequently, hydrogen embrittlement is generally considered to be the mechanism by which corrosion results in cracking, with crack/pit initiation being assisted by localized anodic dissolution. Often, hydrides are not directly observed, but dissolved solute hydrogen is presumed to be present as a result of formation of corrosion products, such as various alloy chlorides. Hydrogen enhanced localized plasticity (HELP), the proposal that elevated solute hydrogen could promote dislocation motion in the vicinity of crack tips \cite{shih1988hydrogen,birnbaum1994hydrogen}, has been suggested as the mechanism by which crack advance occurs, with a characteristic transgranular fracture appearance in Ti-6246 after NaCl HSSCC attack \cite{chapman2015environmentally}. Alternatively or in addition to this, others suggest that adsorption induced dislocation emission (AIDE) could be also operating during stress corrosion cracking (SCC) \cite{lynch1988environmentally}, \textcolor{black}{which refers that the adsorption of hydrogen atoms into subsurface can facilitate dislocation emission ahead of crack tip by weakening the interatomic bonds.} This proposal is supported by TEM observations of high dislocation density below primary $\alpha$ SCC facets in Ti-8Al-1Mo-1V \cite{cao2017mechanism}.

Unlike fingerprint or seawater-associated chloride cracking, HSSCC associated with AgCl has been examined relatively rarely. It was first reported in 1966 where a Ti-7Al-4Mo compressor disc crack was found in a region in intimate contact with silver-plated bolts, leading to failure of the disc in a spin test \cite{duttweiler1966investigation}. X-ray diffraction identified the presence of AgCl at the crack origin. Silver is often used as an anti-galling coating on fixtures such as nuts and bolts. It was believed that AgCl can be generated by the reaction of Ag with trace, ppm-level amounts of chlorine present in the atmosphere, e.g. in solvents, triggering HSSCC above \SI{300}{\celsius}. It continues to be crucial to understand these cracking mechanisms to allow the safe design and operation of gas turbine Ti components.  However, the reaction sequence and mechanism of cracking for the AgCl case have not been examined in detail, with studies being limited to trial-and-error testing and fracture observation.



In this paper, the reaction products associated with AgCl HSSCC in Ti-6246 are analysed. Detailed examination using XRD, SEM and TEM is made of the composition analysis for corrosion products generated both on the surface and within the crack, in order to establish the sequence of reactions in the mechanism chains that gives rise to cracking. This work gives further insight into which alloying effects are most detrimental, and also temperature and thermodynamic effects. These findings will assist in future alloy selection and development, understanding of the interaction with the operating environment (e.g. temperature), and also anti-galling coating selection to avoid occurrence of the problem.

\section{Experimental Description}
\label{experimental}

\begin{figure}[b].  
   \centering
   \includegraphics[width=60mm]{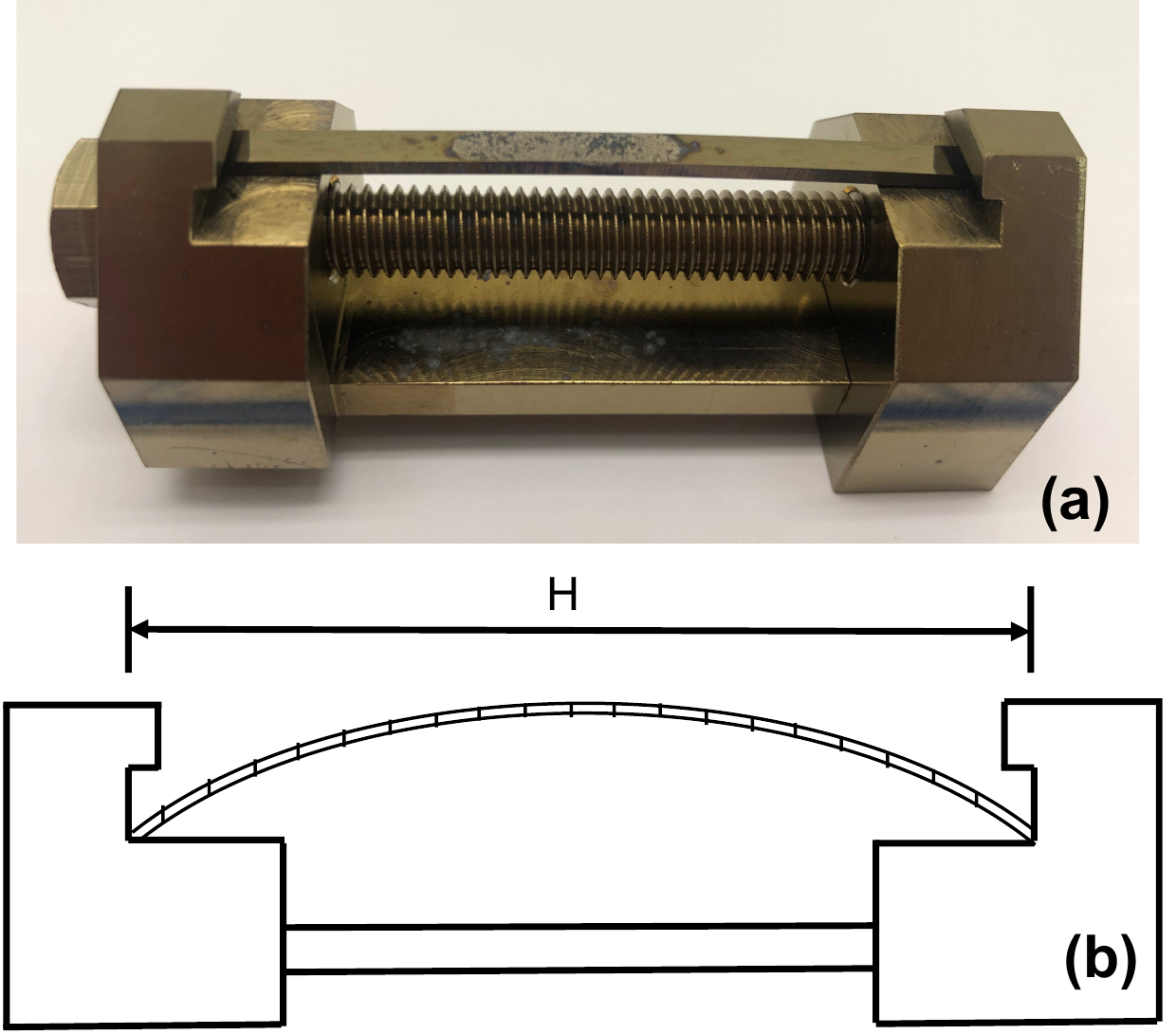}
   \caption{(a) Two-point bend rig used in this work; (b) schematic diagram of the sample and bend rig configuration.}
   \label{fig:bending-rig}
\end{figure}

\noindent Ti-6246 in a service-representative condition was provided by Rolls-Royce plc. $\SI{60}{\milli\meter}\times\SI{1.5}{\milli\meter}\times\SI{3.5}{\milli\meter}$ strip specimens for bend testing were obtained by electrical discharge machining (EDM) and then polished using an OPS colloidal silica solution (1:4). A two-point bending rig was used to apply uniform load across the strip surface, as illustrated in Figure~\ref{fig:bending-rig}. The rig is manufactured from titanium alloys to avoid effects of differential thermal expansion at elevated temperatures on the applied load.  \SI{3}{\milli\gram} AgCl powder (max. particle size \SI{150}{\micro\metre}) was placed on the centre of the strip. A drop of distilled water was added to disperse the AgCl particles before evaporation. Two-point bending tests were then conducted at \SI{380}{\celsius} and \textcolor{black}{500 MPa} for \SI{24}{\hour} in an air-circulating laboratory oven. \textcolor{black}{These are the same conditions as in our previous work, where we examined the sample surface~\cite{shi2020hot}.} The applied stress $\sigma$ was calculated according to ASTM-G41\cite{Solution2011}, as described in following equation
\begin{equation}
H= \frac{K t E(T)}{\sigma}\sin{\frac{L\sigma}{KtE(T)}}
\end{equation}
where $H$ is the distance between two ends of the bending specimen, $K=1.28$ is a constant, $t$ and $L$ are the thickness and length of the specimen, and $E(T)$ is the Young's modulus of Ti-6246 at temperature $T$. The dimensions of the specimen and span length H were measured with a digital caliper ($\pm$0.1mm).

\begin{figure}[t]
   \centering
   \includegraphics[width=95mm]{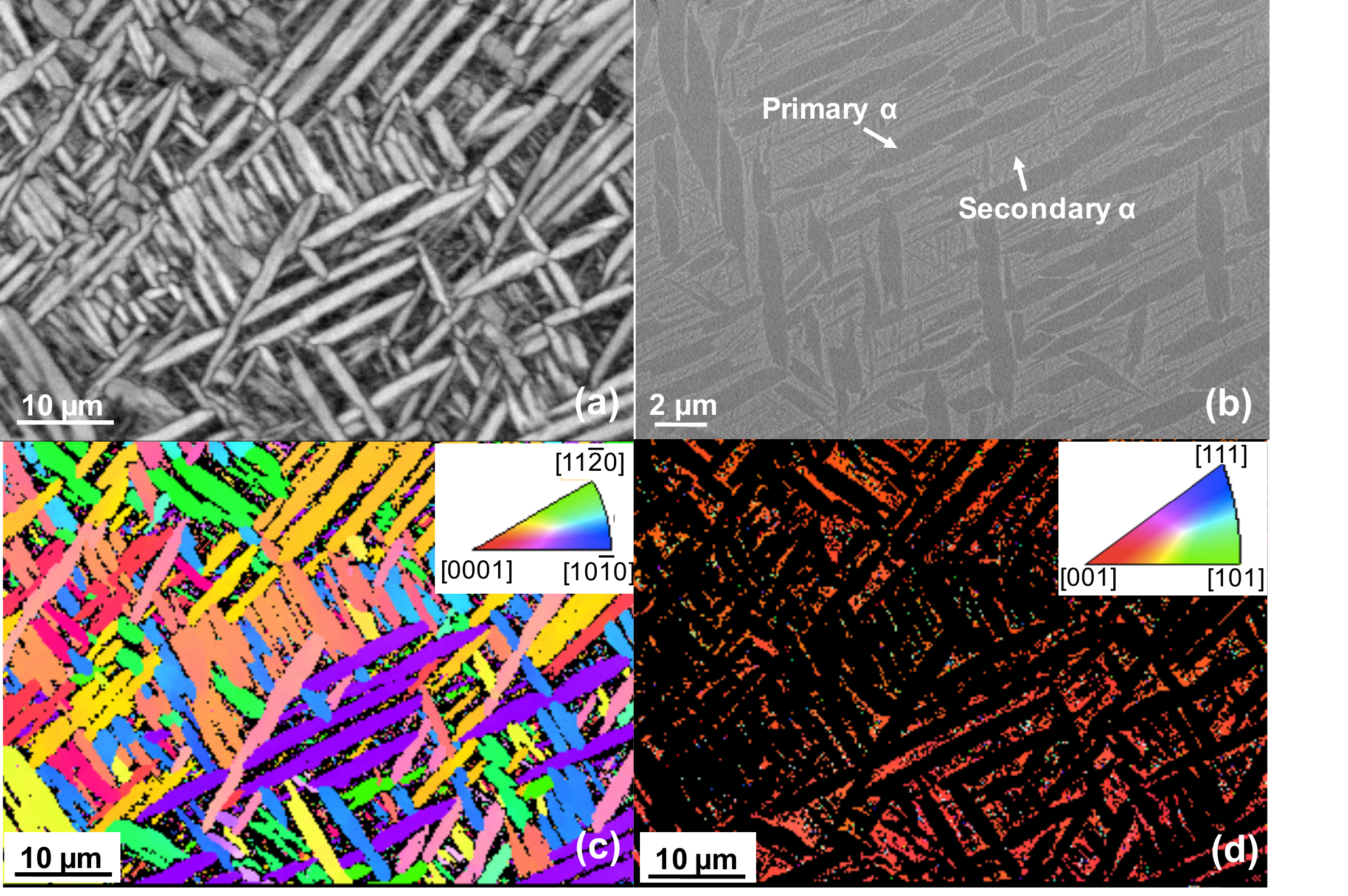}
   \caption{Micrographs showing the microstructure of as-received Ti-6246. (a) band contrast image formed from EBSD; (b) backscattered electron image demonstrating primary $\alpha$ laths and fine secondary $\alpha$ precipitates embedded in transformed $\beta$ phase; inverse pole figure (IPF) maps of the (c) $\alpha$ and (d) $\beta$ phases, coloured according to the loading direction (hoop direction in the parent disc forging).}
   \label{fig:ebsd}.
\end{figure}

The initial microstructure of as-received Ti-6246 was examined by backscattered electron imaging (BEI) and electron backscatter diffraction (EBSD) in a Zeiss Sigma 300 field emission gun scanning electron microscope (FEG-SEM) with an accelerating voltage of 20 kV and a working distance of \SI{15}{\milli\meter}, as shown in Figure~\ref{fig:ebsd}. The specimen surface after corrosion tests was characterised by optical microscopy and with a Zeiss Auriga FEG-SEM under an accelerating voltage of 5 kV.  The cross section underneath the corrosion products was revealed by focused-ion beam (FIB) milling with a Ga$^+$ source at 30 kV and the chemical composition was analysed by energy dispersive X-ray spectroscopy (EDX) equipped in the same SEM. The accelerating voltage was 10 kV and the aperture size was \SI{60}{\micro\metre} for this analysis.  
A cracked sample was characterised by transmission electron microscopy (TEM) and scanning TEM (STEM) in a JEOL-2100F at 200 kV. A TEM foil containing a crack was lifted out from the top surface with a dual beam FEI Helios Nanolab 600. A platinum protective layer was deposited on the surface to minimise damage from ion milling. The foil was made electron transparent with a thickness around \SI{150}{\nano\metre}. Chemical analysis was then carried out using STEM-EDX with a 1 nm spot size. X-ray diffraction(XRD) was also conducted on the sample top surface to investigate the phases formed after corrosion testing, using a Cu K$\alpha$ \SI{1.54}{\angstrom} source with step size \SI{0.03}{\degree}/s. 

\section{Results}
\label{results}
\subsection{Surface observations for tested samples}

\begin{figure}[t]
   \centering
   \includegraphics[width=80mm]{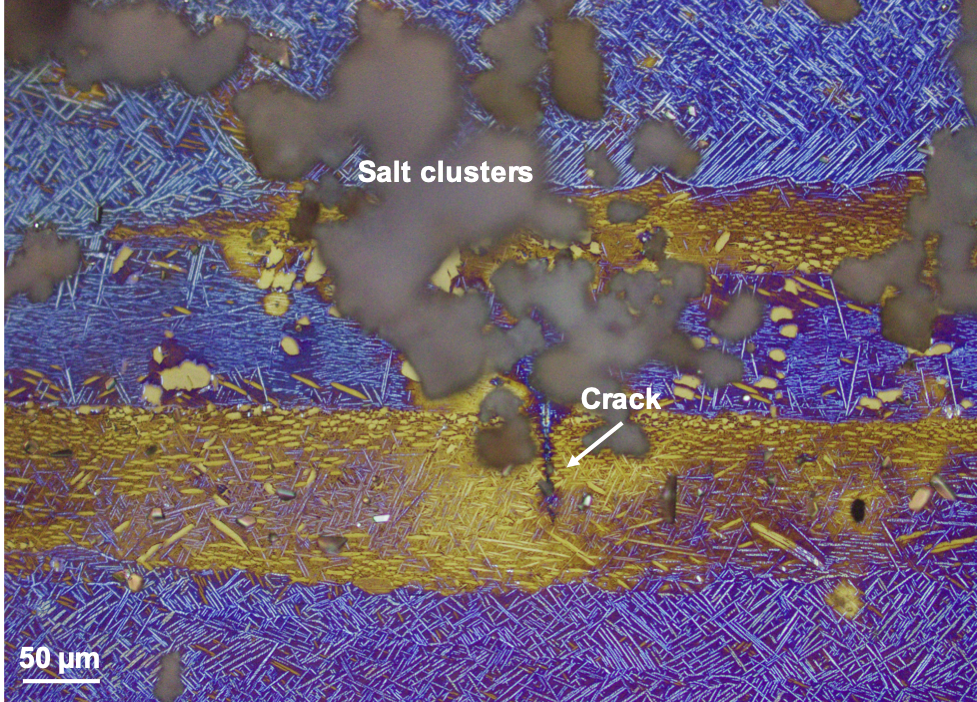}
   \caption{Optical micrograph of the top surface of sample tested under \SI{500}{MPa}/\SI{380}{\celsius} after isothermal heat exposure for \SI{24}{\hour} \textcolor{black}{(image from our previous paper \cite{shi2020hot}}, taken in normal light. A small crack observed is highlighted with a white arrow, as are the AgCl particle clusters on the surface.}
   \label{fig:optical}
\end{figure}

The macroscopic appearance of the sample surface after exposure to AgCl stress corrosion testing  at \SI{380}{\celsius} under a stress of 500 MPa for \SI{24}{\hour} is shown in Figure~\ref{fig:optical}. It can be observed that an uneven colouration appeared on the surface, corresponding to light interference with the oxide film formed during testing. Thus the colour relates to the oxide thickness\cite{delplancke1982self,van2004colour}, which is dependent on the $\alpha$ orientation and local oxidation/corrosion conditions. Such heat tinting is often used in weld qualification, anodization or as a diagnostic tool for thermal exposures. A straw-yellow colouration was observed in regions far from AgCl salt exposure, often with thicker (blue) oxide scales closer to the salt exposure. The underlying microstructure also affected the colours observed, highlighting different prior beta grains. A small crack was found, indicated by a white arrow in Figure~\ref{fig:optical}, on top of which reaction products could be found.

\textcolor{black}{In Figure 3, the prior $\beta$ grains are visible through variations in colour, most likely reflecting the oxide film thickness on the primary $\alpha$ laths and their orientations.  Turning to Figure 4, it appears that the primary $\alpha$ laths are depressed relative to the transformed $\beta$ regions, i.e. that the $\alpha$ phase corrodes more. However, FIB sectioning (Figures 5 and 6) does not completely bear this out. From a fundamental perspective, the $\alpha$ phase contains more Al, which easily forms volatile chlorides, but the $\beta$ phase has far higher solubility for hydrogen.}

\subsection{SEM-EDX}

\begin{figure}[t]
   \centering
   \includegraphics[width=80mm]{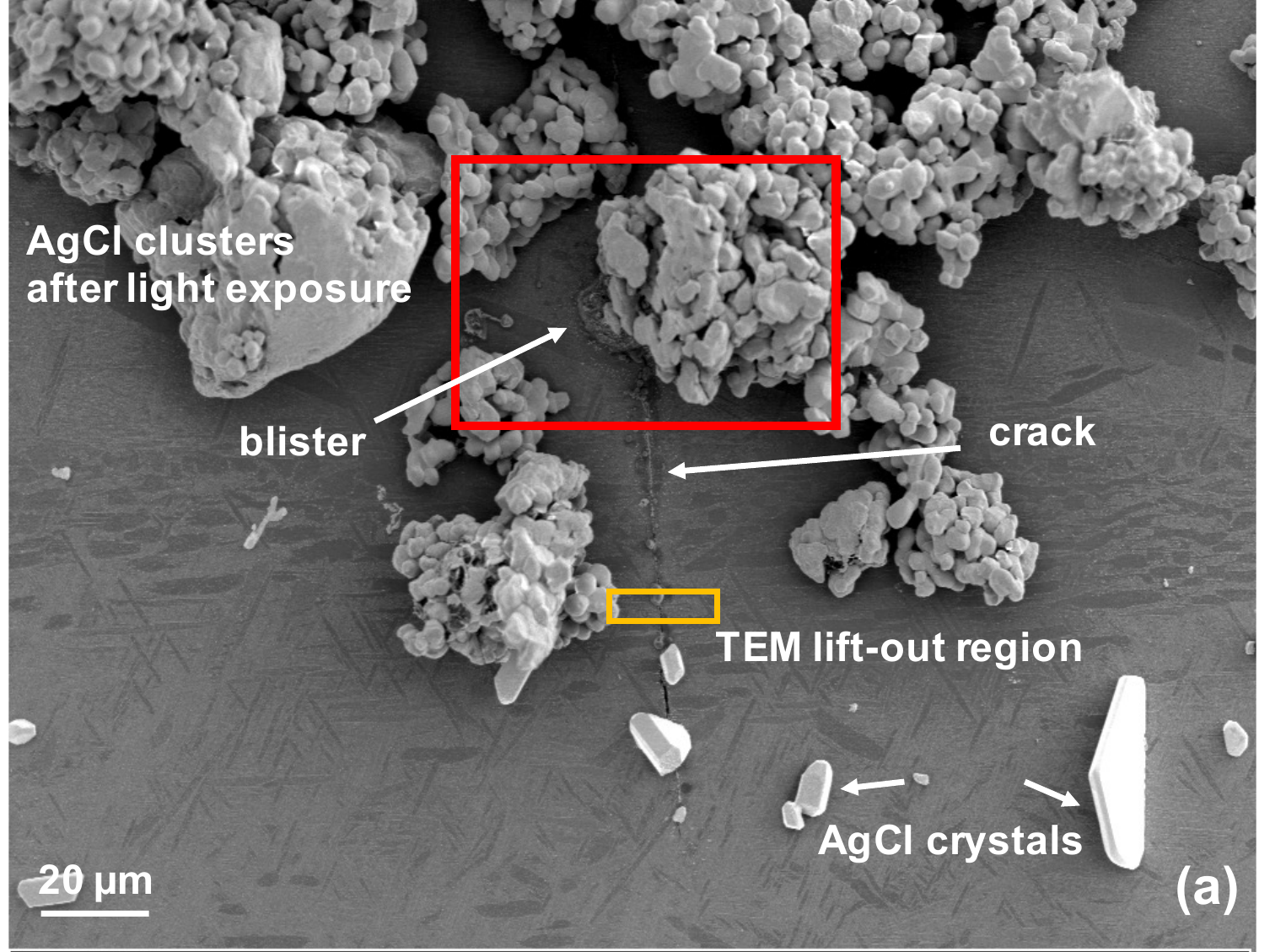}
   \label{figure:SE1}
   \vfill
   \includegraphics[width=82mm]{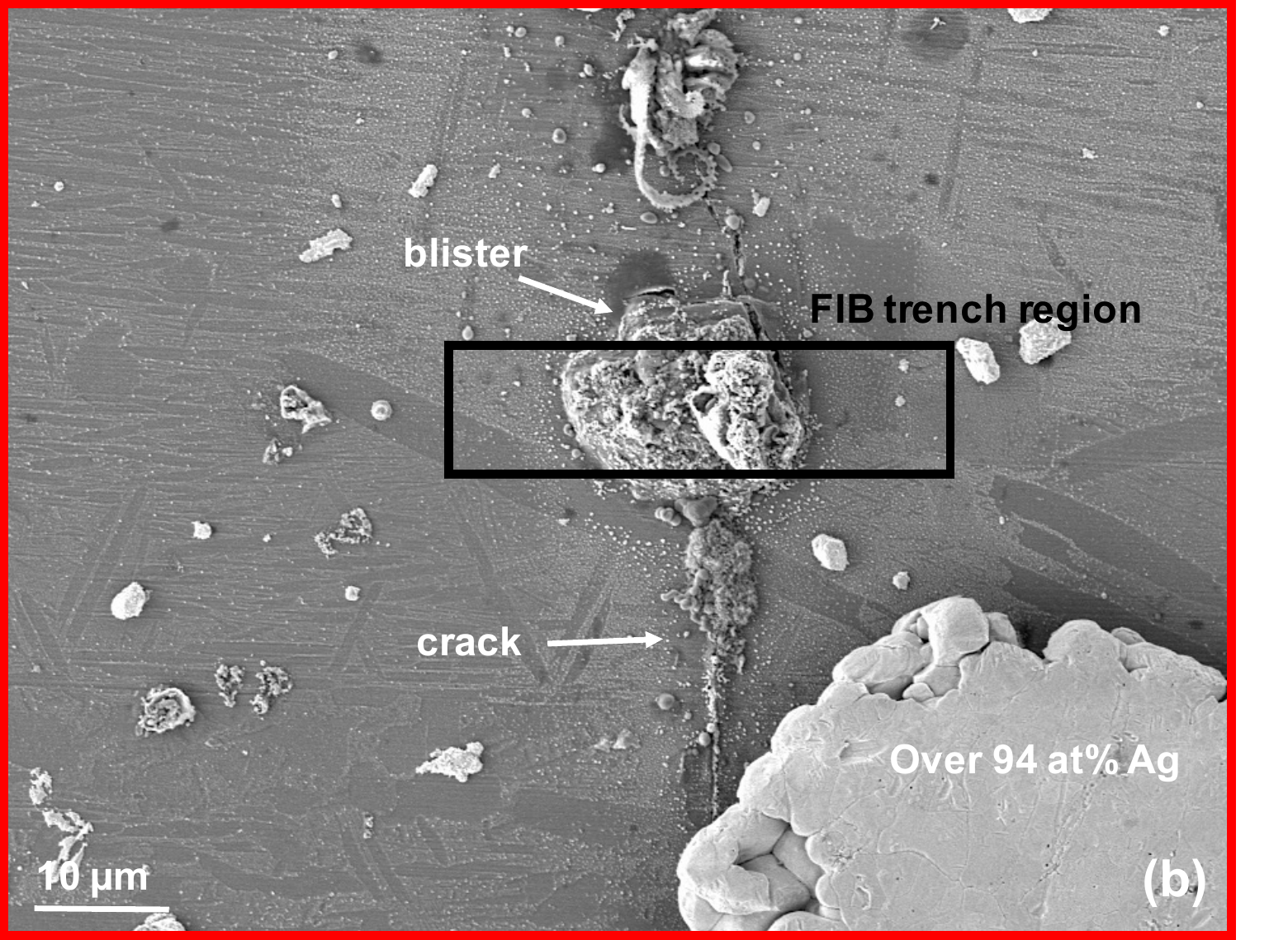}
   \label{figure:SE2}
   \vfill
   \caption{Secondary electron images of the tested sample surface after exposure at \SI{500}{MPa}/\SI{380}{\celsius} for \SI{24}{\hour}, showing a crack near salt clusters. (a) A blister formed underneath a salt cluster and (b) highlighted region from (a) exposed after removal of salt overburden by ultrasonic cleaning. The black box indicates the region protected by platinum coating prior to FIB milling to produce a trench for cross-sectional analysis. A TEM foil was lifted out away from the blister, as outlined by the orange box in (a).}
   \label{fig:SEM1} 
\end{figure}

 The SEM image in Figure~\ref{fig:SEM1} illustrates a small surface crack adjacent to some clusters of AgCl. \textcolor{black}{Unlike the granular and lenticular AgCl crystals as indicated by the bottom two white arrows in Figure~\ref{fig:SEM1}(a), the salt clusters near the crack were composed of many nodular globular particles embedded in larger ones.} EDX analysis on the area labeled in Figure~\ref{fig:SEM1}(b) indicated the presence of more than 94 at\% of Ag, which implied formation of metallic Ag on the applied AgCl. These features have been observed in previous work \cite{lin2013atmospheric} and it is widely agreed that AgCl can photodecompose into metallic silver and chlorine under UV radiation, causing surface or volume darkening of AgCl \cite{moser1959optical,hamilton1988silver}, as follows
\begin{align}
\ce {2AgCl + h\nu &-> 2Ag + Cl2}
\end{align}

Blisters and globular corrosion products were then exposed above the crack after removal of large clusters by ultrasonic cleaning in a dry beaker without any solvents, as shown in Figure~\ref{fig:SEM1}(b). A region of a blister was protected by platinum coating then focused-ion beam (FIB) milled, as indicated by the black rectangular box, to reveal the cross section and conduct chemical analysis shown in Figure 5(a). It can be seen that layered and porous corrosion products were generated inside the blister. They were found to accumulate into a shallow pit, also filling the crack propagating below. This crack began by propagating through a primary $\alpha$ grain exposed at the surface. 
\begin{figure}[t!]
   \centering
   \includegraphics[width=75mm]{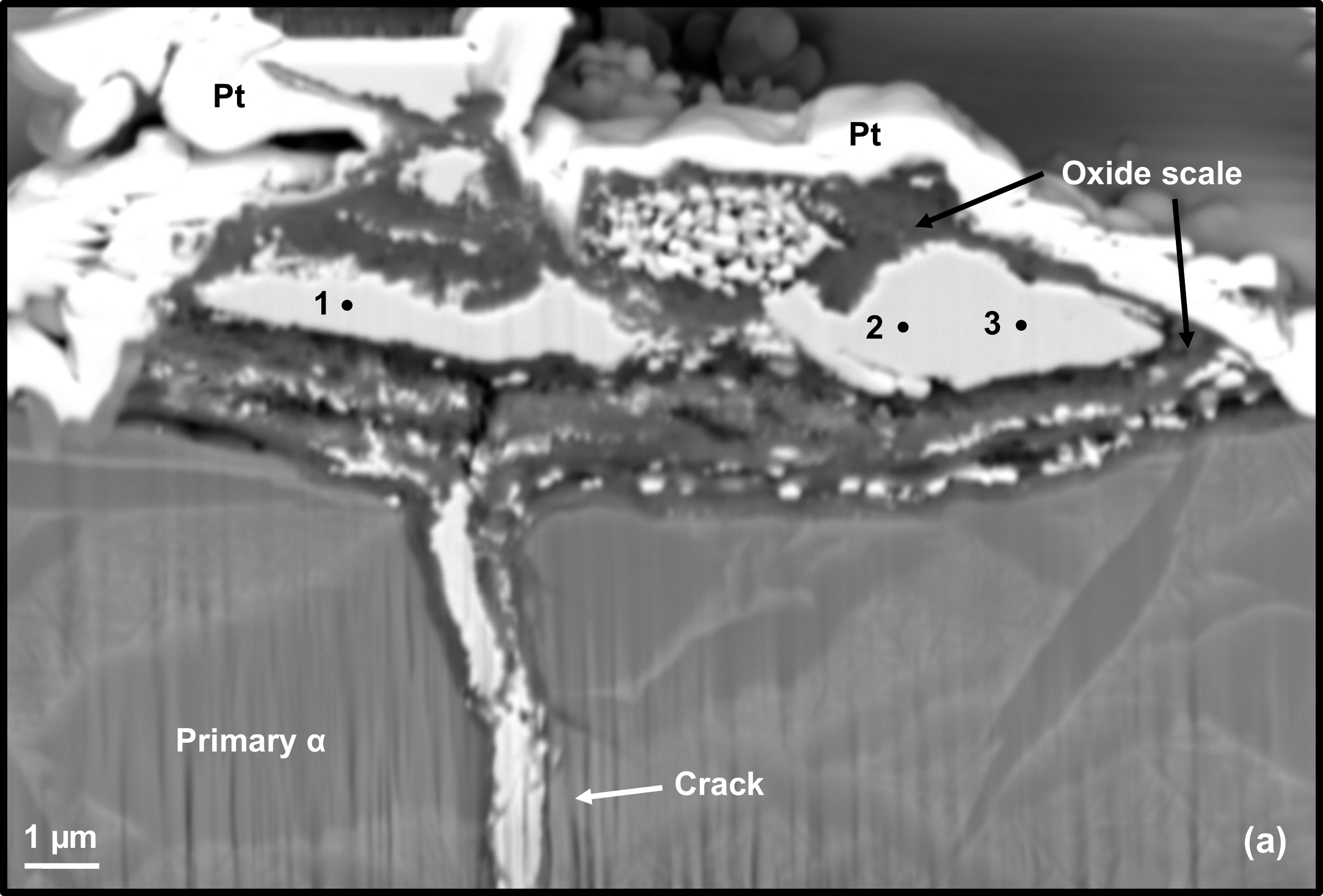}
   \label{fig:SEM}
   \vfill
   \includegraphics[width=90mm]{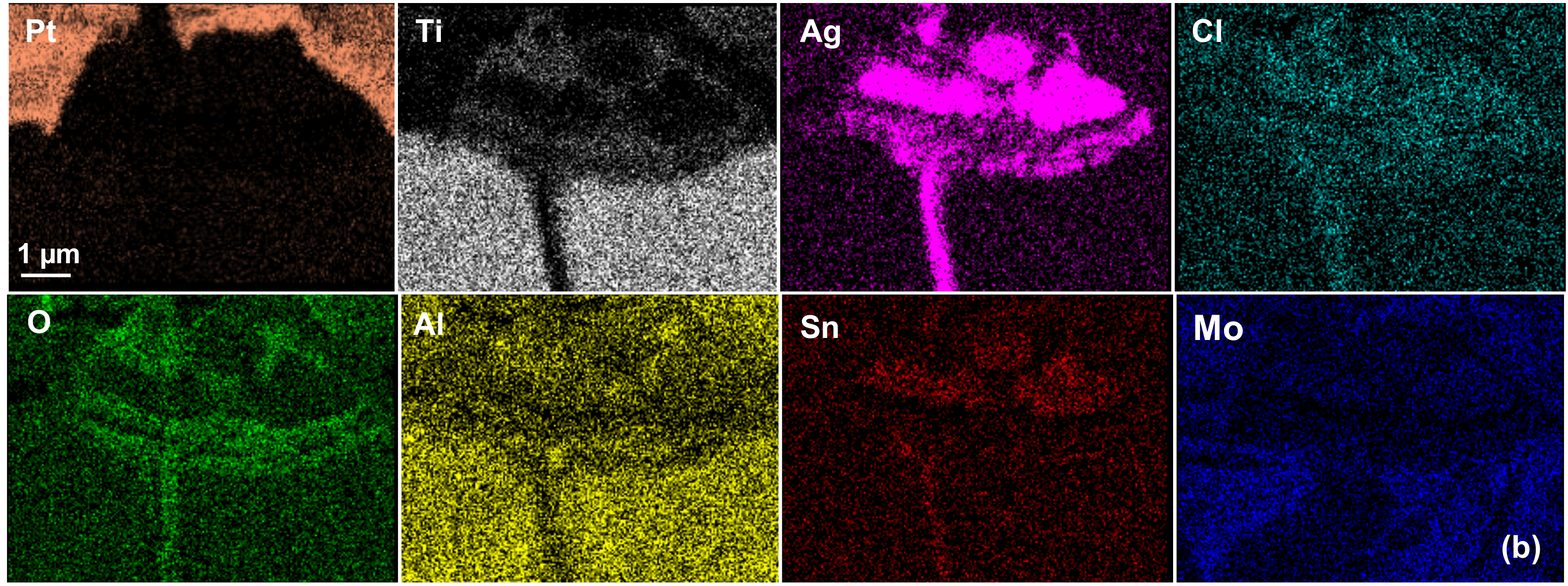}
   \label{fig:EDX}
   \caption{(a) Backscattered electron image revealing the underlying layered structure inside the blister (Figure 4b) after FIB milling; (b) SEM-EDX elemental maps of the red highlighted region}
   \label{SEM-EDX}
\end{figure}

 EDX point analysis in Table 1 suggested that the bright Ag-rich layers with different morphology were surrounded by (grey) oxide scales enriched in Ti, Al and O, e.g. consisting of titanium and/or aluminium oxides.  This implies that Ti and Al were likely active and diffused outward during the corrosion reactions. The top layer of oxides showed a heterogeneous appearance with inclusion of Ag-rich particles, moving inward though the blister. The top oxide/Ag adherence was intimate whereas oxide spallation was observed deeper in the blister, but above the base alloy.  Cellular and globular Ag-rich structures were generated on the top, surrounded by a dense oxide layer around \SI{1}{\micro\meter} in thickness within the blister. Aggregation and coalescence of Ag-rich particles were also observed at the bottom of the pit. It has been suggested that local chemical potential might control the Ag formation morphology, which was closely linked to reaction kinetics. It is also well known that Ag prefers island growth \cite{kulczyk2014investigation} (The mechanism of Ag/oxide alternation is unknown).

Point analysis of this dense Ag-rich layer, Table 1, demonstrated that it contained approximately 80 at.\% Ag with Ti (13-15 at.\%) and small amounts of Al or O. In the binary Ag-Ti system, Ag has relatively small solubility for Ti($<$ 5 at.\%) at \SI{380}{\celsius} such that an intermetallic phase of TiAg is preferably formed \cite{murray1983ag}. Such solubility as exists would be increased by the presence of Al\textcolor{black}{, at least at high temperatures~\cite{Lukas1990ag}}; in contrast the Ag-Ti-O ternary phase diagram has not yet been assessed. The backscattered electron image in Fig 5(a) show some fine-scale speckled contrast below the spatial resolution of SEM-EDX which may indicate the formation of intermetallic TiAl in the silver layer, which would be consistent with the Ag-Ti phase diagram. However, it is difficult to examine composition variations within this layer in more detail without resorting to EPMA or STEM-EDX, as the resolution of SEM-EDX in these imaging conditions is around \SI{1}{\micro\meter}. It should also be recognised that the main Sn (K$_\alpha$ 3.44 keV ) and Cl (K$_\alpha$ 2.62 keV) peaks overlap with the minor Ag L$_{\beta 2}$, L$_{\beta 10}$ and L$_1$ peaks at 3.35, 3.43 and 2.63 keV, respectively. Therefore, as the Sn and Cl maps spatially match with locations where Ag is present, it is less possible to definitively conclude that small trace amounts of Sn and Cl are present using SEM-EDX. Likewise, the compositions of thin and discontinuous oxide layers could not be accurately examined by point analysis under this condition due to the limited spatial resolution.

\begin{table}[b!]
\centering
\caption{SEM-EDX point analysis of location in the Ag-rich layer in Figure 5(a).}
\begin{small}\begin{tabular}{c c c c}
    \hline
     at.\%&Point 1&Point 2&Point 3  \\
     \hline
     Ag& 83.9&79.2&82.4\\
     Ti& 13.6&15.5&13.3\\
      O& 0    &5.4 &  4.3\\
     Al &2.5 &0    &  0   \\
     \hline
\label{table1}
\end{tabular}\end{small}
\end{table}

\subsection{STEM-EDX analysis}

A TEM foil was lifted out from the top surface of the sample from the location highlighted in Figure 4(a) after a test where a crack propagated through the surface, which was covered by small blister-like corrosion deposits. STEM bright-field (BF) imaging of the foil, Figure 6(a) revealed that the crack penetrated into the underlying alloy. \textcolor{black}{The crack path propagated through the $\alpha$ laths, with branching in between, with the crack itself observed to be partially filled with heterogeneous corrosion products. Cracking at the interface between the primary $\alpha$ and the transformed $\beta$, \emph{e.g.} due to hydride formation at the interface, was not observed.}

Regions of bright contrast in Figure 6(a) but with no corresponding intensity in the STEM-EDX maps, Figure 6(b) are inferred to be empty regions of the crack. Regions of the crack of low intensity (black) in the BF image were found to be associated with high contents of Ag.  The crack wall itself was found to be elevated in O content.

\textcolor{black}{In addition, there is found to be a co-occurence of Sn with Ag in the SEM-EDX maps. Inspection of the X-ray spectra shows that these peaks are clearly distinct, and the Ag-Sn phase diagram \cite{karakaya1987ag} suggests that Ag has substanital solubility for Sn. Sn chlorides have quite low evaporation points and so it is implied that Sn is involved in the corrosion process, generating e.g. gaseous SnCl$_4$ which then are oxidised to SnO$_2$ and deposited on the Ag alloy particles.}

The small particles inside the crack were often found bound together into clusters over \SI{200}{\nm} in size. These large particles were examined by point analysis in STEM-EDX, smaller than the particle size, as shown in Table 2. Here, much higher O contents were found, in the region of 35-45 at.\%, along with Ag, Ti and the other metals in Ti-6246, i.e. Al, Mo, Sn and Zr.  Since Ag does not form stable oxides at the test temperature of \SI{380}{\celsius} \cite{ellingham1944reducibility}, it is suggested that the regions of the particles sampled by STEM-EDX are composed of mixtures of silver and Ti-based metal oxides. However, the O:Ti ratios do not clearly indicate the presence of a particular oxide stoichiometry, such as TiO$_2$.  In addition, quite significant amounts of Cl were found, with the amount increasing to nearly 6 at.\% with depth along the crack, which might suggest the increasing difficulty of evaporation of chlorides with distance from the crack mouth. In general, the amounts of Al relative to Ti are higher than the alloy content, possibly reflecting the higher stability of Al chlorides formed than Ti chlorides, whereas the converse is true for Sn \cite{cottrell2019introduction}. Those chlorides can then react to form oxides.


\begin{figure}[t!]
   \centering
   \includegraphics[width=70mm]{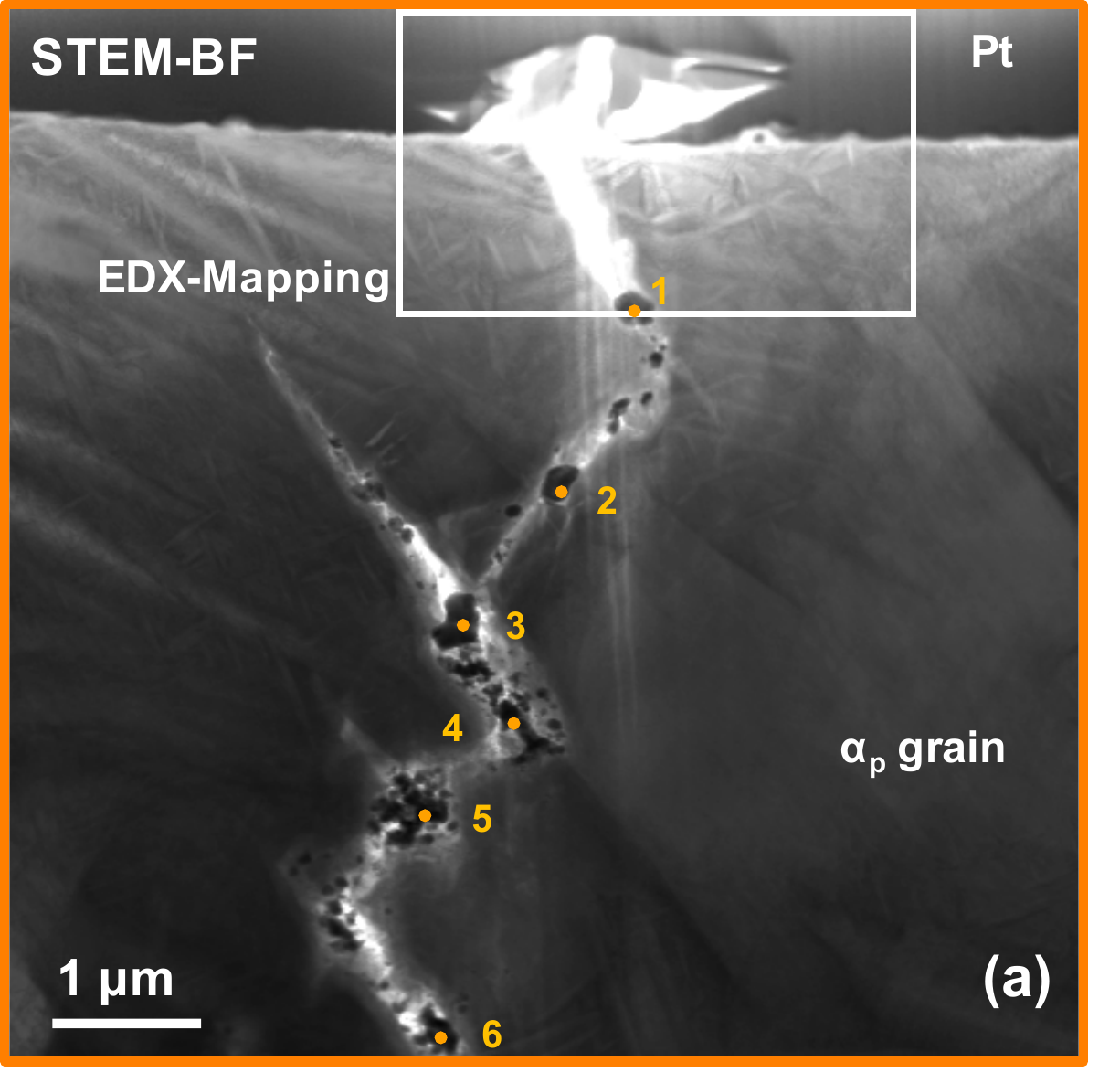}
   \label{fig:TEM foil}
   \includegraphics[width=90mm]{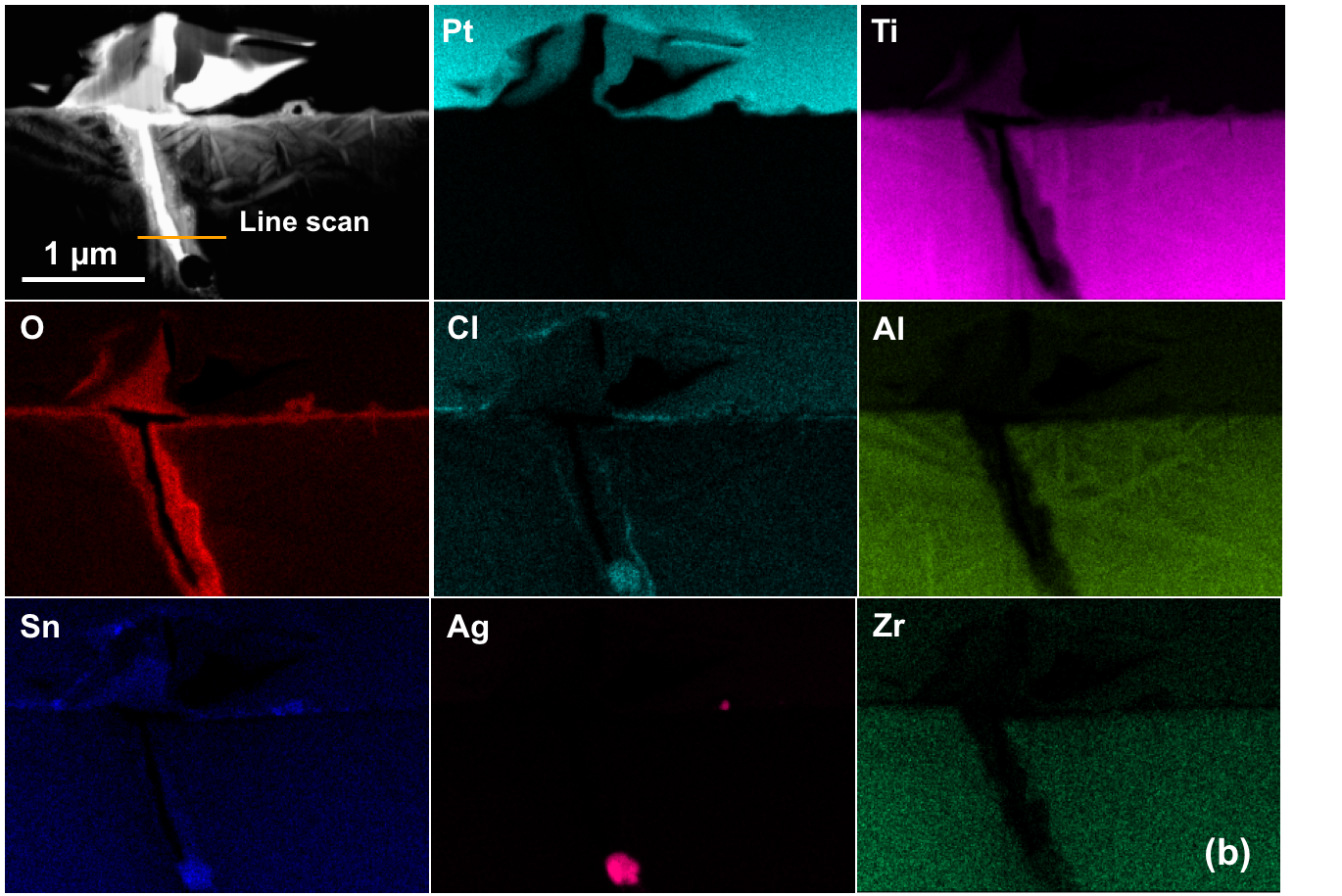}
   \label{fig:STEM-EDX}
   \includegraphics[width=65mm]{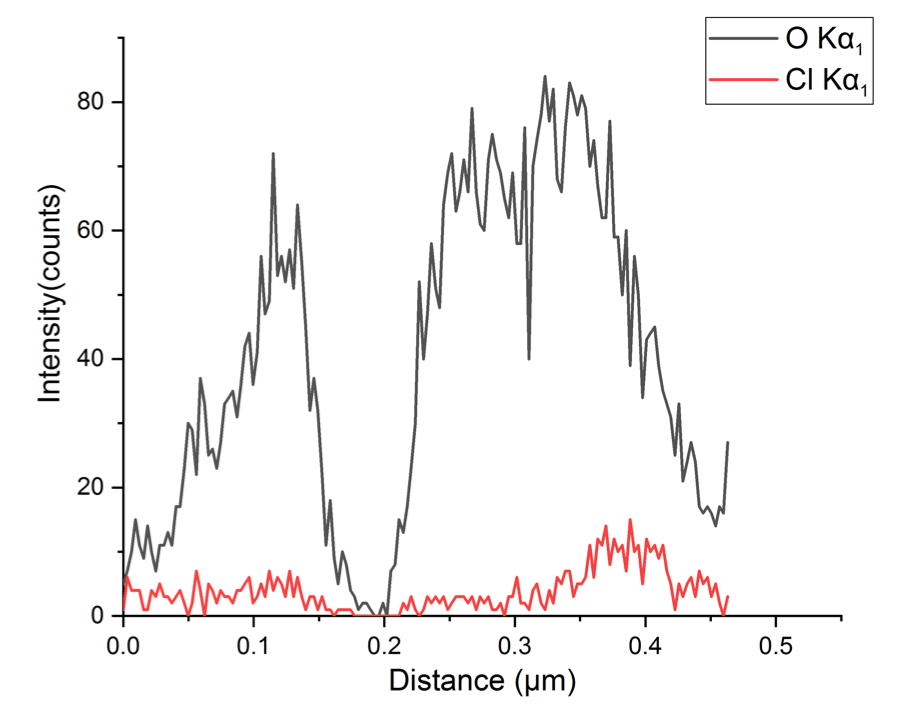}
   \caption{(a) Bright-field STEM image (spot size=1nm) of the TEM foil lifted out from the crack in Figure 4(a)\textcolor{black}{, showing a transgranular crack propagation path}; (b) STEM-EDX elemental maps of the region highlighted in (a), and (c) STEM-EDX line scan analysis showing intensity variations of O and Cl along the line identified in (b). Elevated levels of Cl are present at the metal/oxide interface.}
   \label{STEM-EDX}
\end{figure}

\begin{table}[b!]
\centering
\caption{STEM-EDX point analyses (at.\%) at the points denoted in the Ag-rich layer within the crack in Figure~\ref{STEM-EDX} (Cu removed, e.g. from the sample holder).}
\begin{small}\begin{tabular}{c c c c c c c}
    \hline
    Point&1&2&3&4&5&6 \\
     \hline
     Ag& 44.0&30.5&23.7&36.3&21.4&29.1\\
      O& 39.5&34.2&46.7&31.3&44.6&36.3\\
      Ti& 13.2&29.8&21.2&24.2&24.1&22.8\\
      Al&  1.4& 3.6& 6.7& 3.6& 4.5& 4.9\\
      Cl&0.7&0.9&1.1&2.3&4.3&5.8\\
      Mo&0.4&0.6&0.5&2.1&1.0&1.0\\
      Sn&0.6&0.2&0&0&0&0\\
      Zr&0.1&0.2&0.1&0.2&0.2&0.1\\
     \hline
\label{table2}
\end{tabular}\end{small}
\end{table}

\begin{figure}[t]
   \centering
   \includegraphics[width=80mm]{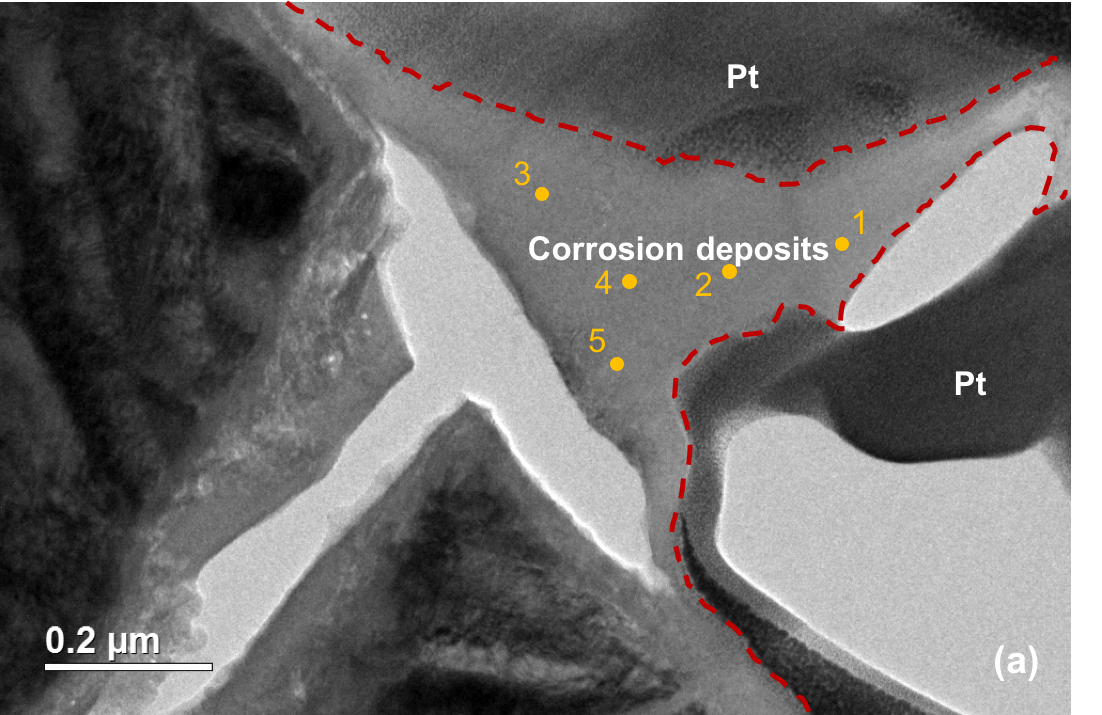}
   \includegraphics[width=80mm]{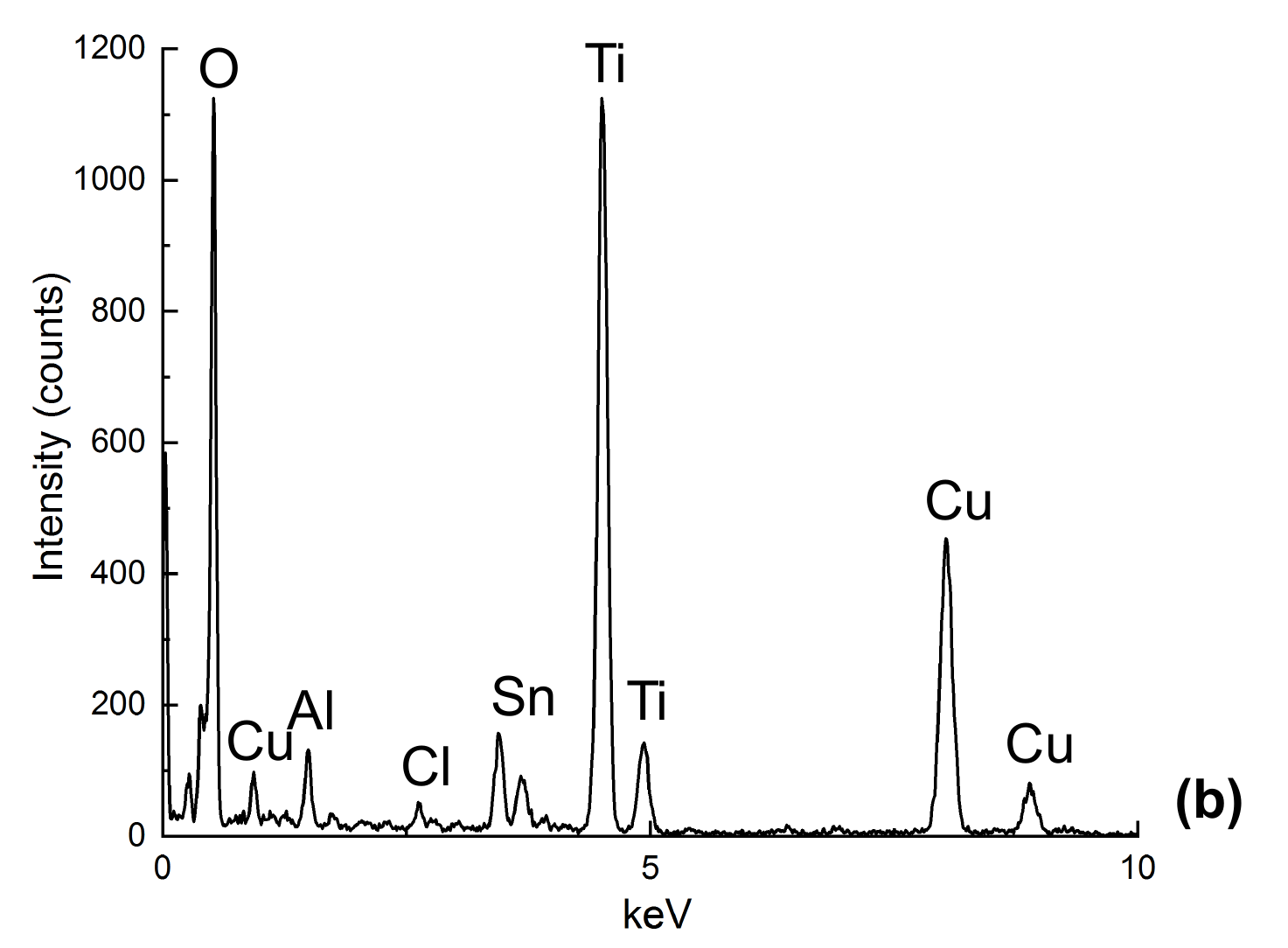}
   \caption{(a) BF TEM image of the corrosion deposit formed above the crack mouth; (b) EDX element spectrum obtained from point analysis (location \#1) of the corrosion deposits;}
   \label{fig:crackmouth}
\end{figure}

\noindent Figure 7(a) shows the crack mouth region highlighted in Figure~\ref{STEM-EDX} in more detail. The underlying $\alpha/\beta$ Ti microstructure contains a high defect density caused by Ga damage from sample preparation. A thin layer of corrosion deposits was found above the crack mouth, which is enriched with oxygen according to the STEM-EDX mapping results in Figure 6(b).   Above these the Pt deposition layer is observed, which is incomplete due to FIB damage.  Point analyses of the corrosion deposits were undertaken (1-5), Figure \ref{fig:crackmouth}(b) and Table \ref{table3}. The precision quoted is 1 at.\% for concentrations $>$10 at.\% and 0.1 at.\% for concentrations $<$10 at.\%. It is inferred that the corrosion deposits are most likely predominantly based on TiO$_2$, but that diffraction based analyses are required to provide confidence as to the phase assignment. The Al:Ti content, at around 1:10, reflects the approximate alloy composition (10.8:83 in at.\%), whilst the Sn content is substantially elevated relative to the content in base alloy (0.8 at.\%). Therefore it seems likely that this reflects the ease of forming tin chlorides with high volatility \cite{gamsjager2012chemical}.  As anticipated, unlike in the SEM-EDX measurements, Ag was not observed in these corrosion deposits. It might be attributed to its location far from salt particles with absence of initial reactions with AgCl.

\begin{table}[t!]
\centering
\caption{STEM-EDX point analysis on the corrosion deposits above crack (Cu was removed).}
\begin{small}\begin{tabular}{c c c c c c }
    \hline
     Element in at.\%&1&2&3&4&5 \\
     \hline
     O& 70&79&69&69&70 \\
     Ti& 22&16 &26&26&26 \\
     Al& 2.8&1.9&2.6&2.3&2.6 \\
     Sn& 3.7&2.4&1.9&2.1&1.6 \\
     Cl& 1.0&0.6&0.5&0.6&0.5  \\
     \hline
\label{table3}
\end{tabular}\end{small}
\end{table}

\subsection{X-ray diffraction analysis}
\begin{figure}[t!]
   \centering
   \includegraphics[width=80mm]{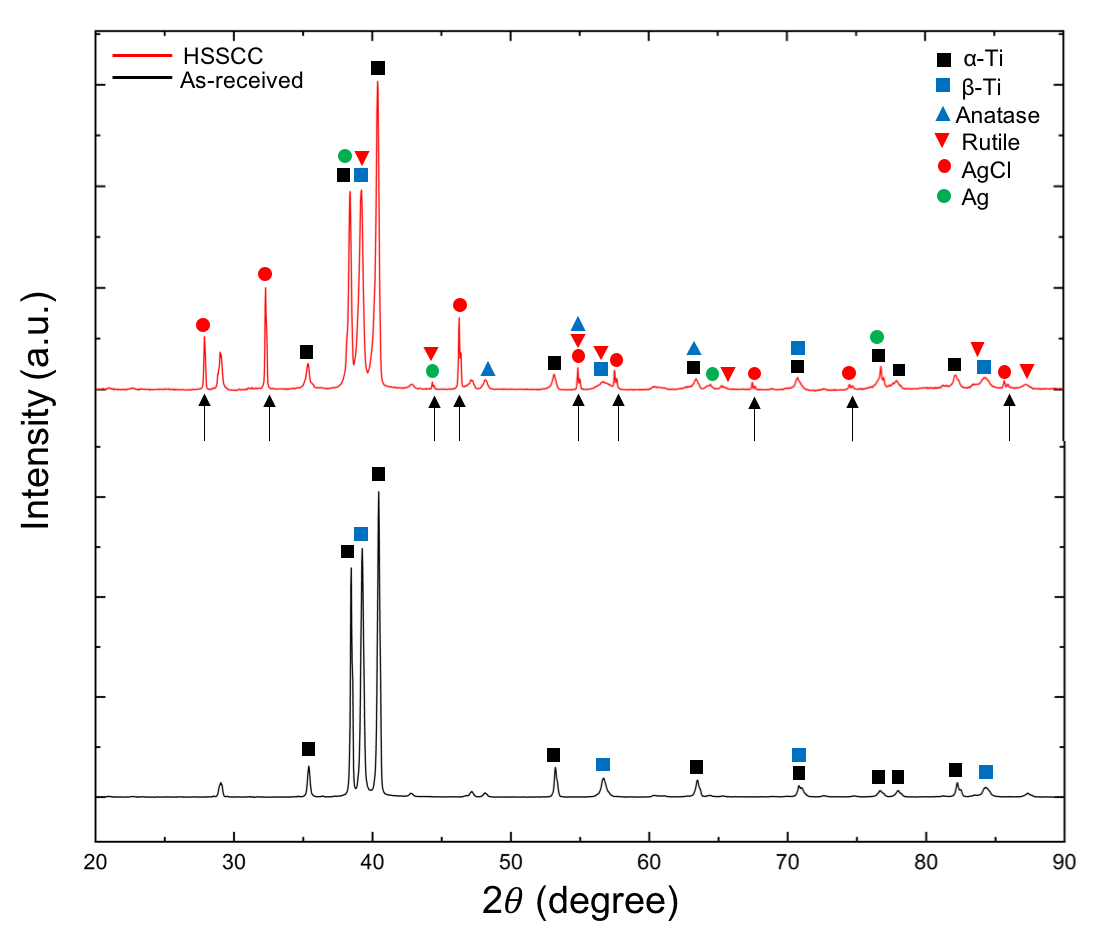}
   \caption{X-ray diffraction pattern obtained (Cu $K_\alpha$ radiation) from the as-polished sample surface and after stress corrosion cracking testing.}
   \label{fig:xrd}
\end{figure}

To complement the EDX results, bulk powder X-ray diffraction was conducted on an as-corroded sample surface that had been tested at \SI{380}{\celsius} for \SI{24}{\hour}, Figure~\ref{fig:xrd}. The primary $\alpha$ and $\beta$ phase peaks can be readily identified by comparison to an uncorroded sample.  A number of very small peaks (e.g. at \SI{29.0}{\degree}, \SI{42.9}{\degree} and \SI{47.2}{\degree}) in the as-received sample correspond to background from the sample holder. In the corroded sample, additional peaks indicated by the black arrows were observed, which match well to AgCl. There are indications of the presence of Ag (e.g. the increase in intensity of the peaks at \SI{77}{\degree}), as distinct from TiO$_2$ (the peak at \SI{44}{\degree} which could be either Ag or TiO$_2$). However, the TiO$_2$ films observed using electron microscopy were very thin, so it is perhaps unsurprising the TiO$_2$ is difficult to unambiguously observe. Therefore bulk X-ray diffraction analysis provided only limited additional information, so instead we turn to TEM-based diffraction phase analysis.



\subsection{HRTEM and dislocation analysis}

The different phases present in the corrosion products shown in Figure \ref{fig:crackmouth} were further analysed by high resolution TEM (HRTEM). STEM-EDX results showed that this region contained elevated concentrations of O, Ti, Al and Sn, suggesting the formation of mixed oxides. Selected area diffraction (SAD) patterns from this region were then matched with the crystal structures of \textcolor{black}{TiO$_2$}, SnO$_2$ and Al$_2$O$_3$. In this case, anatase was determined as the form of TiO$_2$ occurring, which is thermodynamically unexpected~\cite{jamieson1969pressure,2009AmMin..94..236S}. However, it has previously been suggested that the formation of nanocrystalline anatase particles can be favoured owing to its lower surface free energy and rapid crystallization~\cite{banfield1998thermodynamic,hanaor2011review}.  Besides, the d spacing measurement of the most intensive diffraction ring was \SI{3.7}{\angstrom}, slightly larger than d spacing of (011) planes for anatase. It was postulated that doping of other elements in anatase might expand the lattice structure \cite{hanaor2011review}.

STEM-BF imaging, Figure 10, of the large primary $\alpha$ grain in Figure 6(a), was performed to highlight the dislocation structures associated with cracking. Long, straight dislocations of basal and pyramidal trace were observed.

\begin{figure}[h]
   \centering
   \includegraphics[width=82mm]{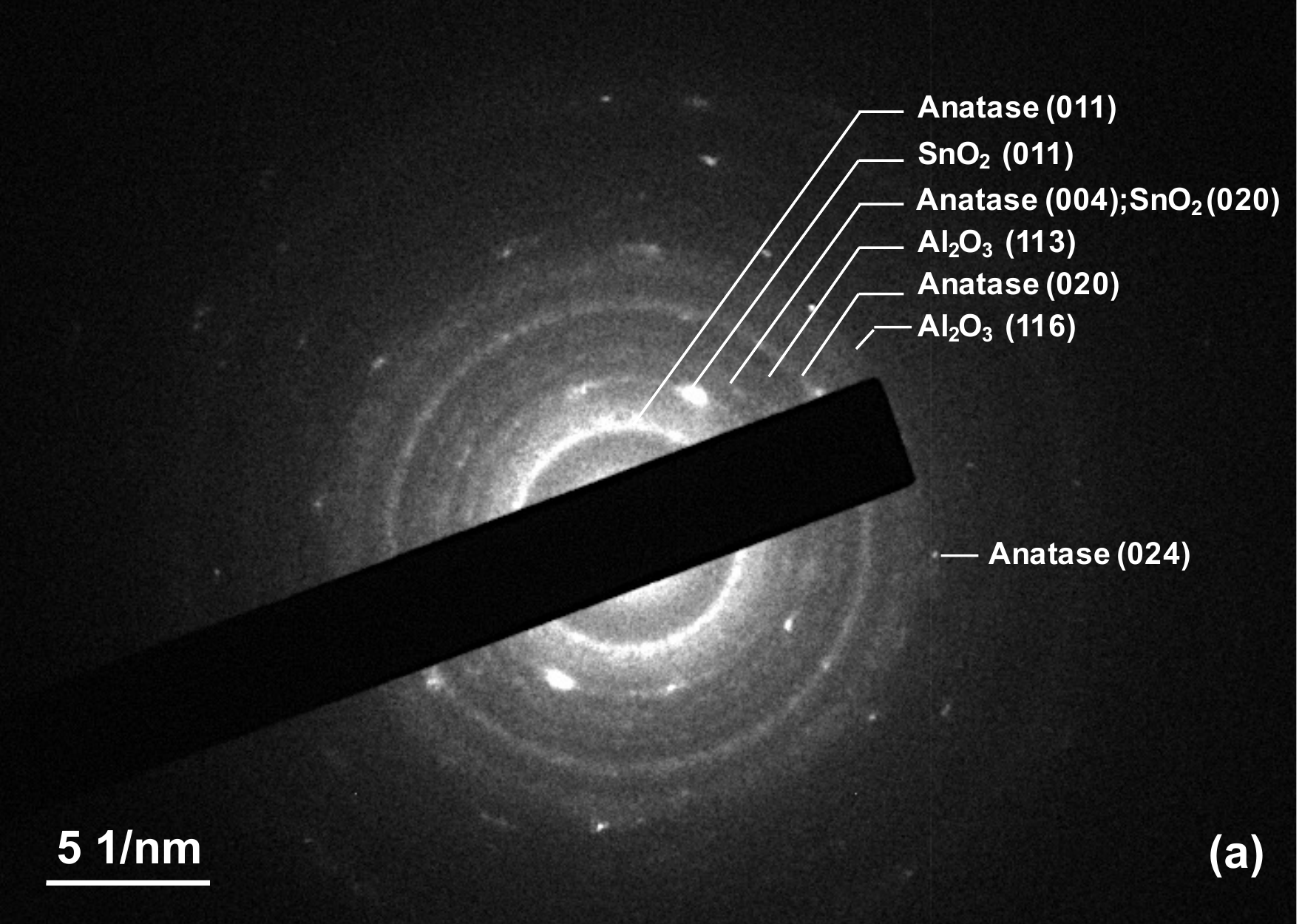}
   \includegraphics[width=82mm]{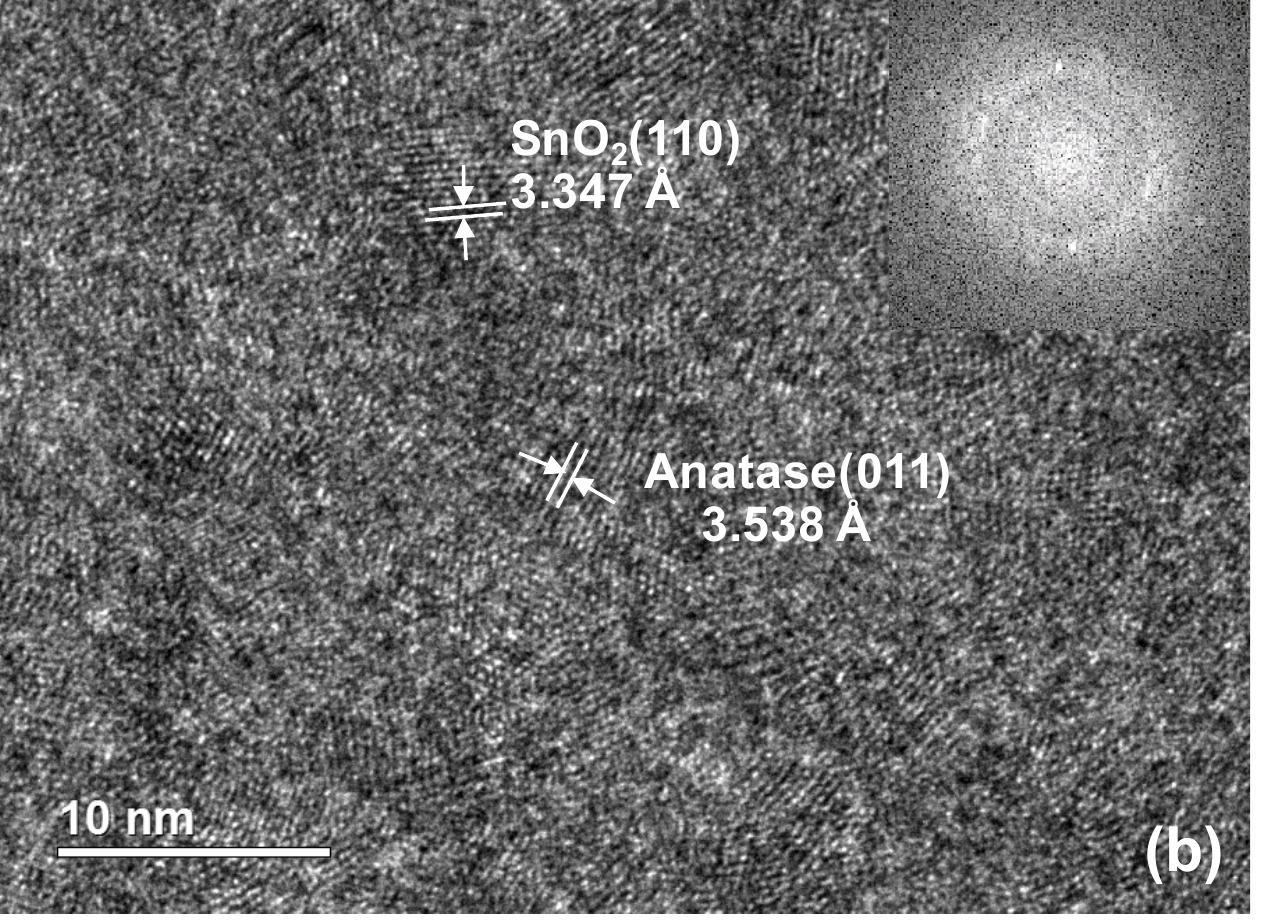}
   \caption{(a) Selected area diffraction (SAD) patterns and (b) high resolution TEM (HRTEM) image of the corrosion products formed above the crack mouth as illustrated in Figure 7(a).}
   \label{fig:HRTEM}
\end{figure}

\begin{figure}[h]
   \centering
   \includegraphics[width=80mm]{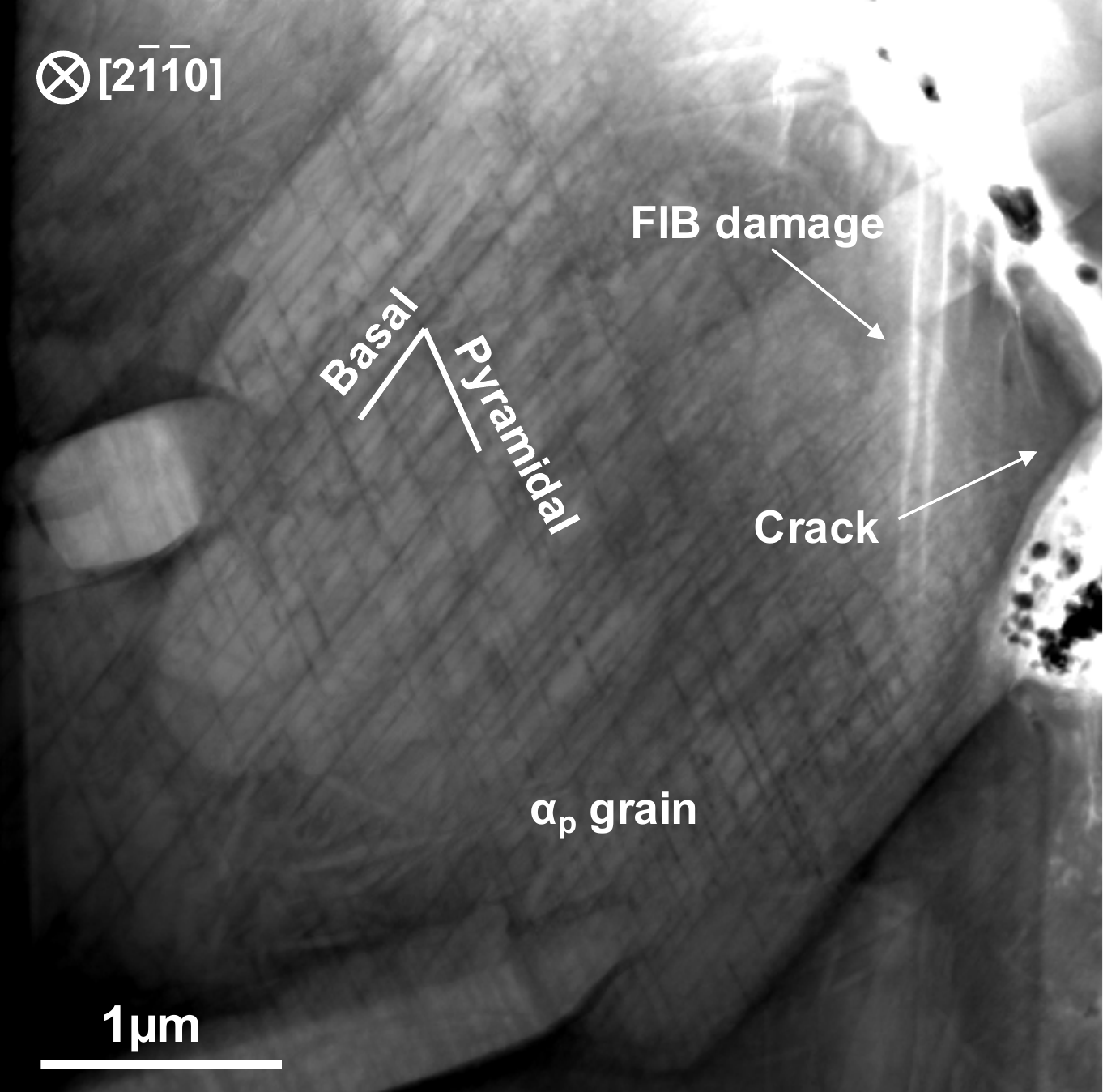}
   \caption{STEM-BF image showing high dislocation density in a primary alpha grain near the crack.}
   \label{fig:Dislocation}
\end{figure}

\section{Discussion}

\subsection{AgCl formation in engines}
The susceptibility of HSSCC induced by AgCl in titanium alloys was firstly reported by Duttweiler on 1966 after failure of a Ti-7Al-4Mo compressor disk during spin testing \cite{duttweiler1966investigation}. Cracking was found at the bolt hole regions in contact with sliver-plated bolts with presence of AgCl deposits. It was suggested that only 0.02 ppm Cl$_2$ in the atmosphere was enough to corrode metallic Ag to produce AgCl salts above \SI{400}{\celsius} as explained in equation (3), which was estimated by the Van't-Hoff reaction. Sources of chlorine can occur in the field under variations in local conditions. Formation of AgCl has been shown to be  energetically favoured over any silver oxides when the bare Ag surface was exposed to Cl$_2$ and OH radicals \cite{liang2010effects,lin2013analysis}, according to
\begin{equation} \mathrm{Cl}_2 + \mathrm{Ag} \rightarrow 2 \mathrm{AgCl} \end{equation}
Duttweiler's results showed that the rupture lives of Ti-5Al-2.5Sn and Ti-7Al-4Mo coated by AgCl were dramatically reduced compared to the bare metal above \SI{370}{\celsius}. The fracture surfaces exhibited brittle features with crack progression through $\alpha$ phase. 
The author also examined the effect of solid state Ag-Ti interactions through metallic coupling. It was found that Ag-plated titanium specimens under vacuum also showed reduced rupture lives, and that the rupture lives of AgCl coated Ti-5Al-2.5Sn specimens can be up to 100 times shorter than that of Ag-plated samples, under higher temperature conditions, \SI{470}{\celsius} and \SI{345}{\mega\pascal}~\cite{duttweiler1966investigation}. Given the age of that report, modern electron microscopy techniques were not available and therefore the mechanism of AgCl-associated embrittlement could only be established by A/B hypothesis testing and fractographic appearance. For example, the possibility of solid metal embrittlement (SME) has not been excluded, and Stoltz and Stuben~\cite{stoltz1978solid} have reported SME of Ti-6Al-6V-2Sn by intimate contact with silver above \SI{203}{\celsius}.  
The likelihood of AgCl formation in engines was considered by Duttweiler to be high and it has previously been established that hot salt attack can have a more deleterious effect than SME in titanium and other metals, and therefore stress corrosion was preferred over solid metal embrittlement as a hypothesized cracking mechanism.

\subsection{Main experimental observations}
In this work, AgCl-induced HSSCC was studied and a Ti-6246 specimen tested under \SI{500}{MPa}/\SI{380}{\celsius} for \SI{24}{\hour} was meticulously investigated. An uneven colouration appeared on the surface after corrosion testing. This phenomenon is also commonly seen when anodizing titanium or after welding or heat treatment \cite{delplancke1982self}. It is inferred that an oxide film of increasing thickness was generated by oxidation enhanced by AgCl-induced corrosion. The oxide film `halo' around each salt particle increased in thickness with proximity to the chloride particle.   Multiple surface cracks were found to initiate underneath salt clusters. Crack propagation followed a transgranular path.  SEM-EDX analysis on the cross-section of a blister formed above the crack mouth revealed the presence of metallic Ag phase and titanium oxides. Significant amounts of Ag-rich phase is observed within the crack, suggesting the occurrence of Ag migration. STEM-EDX examination of the Ag-rich particles within a crack showed that these particles also contained mixtures of different Ti and alloying element metal oxides. 

XRD analysis on the sample surface verified the presence of Ag and AgCl, whilst other corrosion products were below the detection limits of XRD, or shielded by the Ag / AgCl on the surface. TEM study confirmed the presence of TiO$_2$ (in the form of anatase), SnO$_2$ and Al$_2$O$_3$, generated as nanoprecipitates blocking the crack mouth. Moreover, Cl was also observed all the way along the length of the crack and concentrated at the oxide/metal interface, implying the formation of metal chlorides at the testing temperature, \SI{380}{\celsius}. This observation is consistent with findings in previous studies \cite{joseph2018mechanisms,chapman2015environmentally}, which proposed that gaseous HCl could be produced as a byproduct from the reaction of chloride salts with the base metal and atmospheric moisture, and then in turn react with the exposed bare metal at the crack tip, promoting hydrogen charging of the crack tip. Hydrogen embrittlement is hence postulated as the responsible cracking mechanism.

TEM investigation on a brittle-appearance primary $\alpha$ grain displayed a large amount of long and straight dislocation segments on basal and pyramidal planes. This high dislocation density across the whole grain was not observed in previous study on NaCl HSSCC of Ti-6246, especially not adjacent to the fracture surface within \SI{1}{\micro\meter} distance \cite{chapman2016dislocation}. Also, less dislocation entanglements were seen in this case, which could imply a change of the embrittlement mechanism ahead of crack tips for AgCl HSSCC.

\begin{table*}[t!]
\centering
\caption{Calculations of standard formation enthalpy for proposed reactions at \SI{380}{\celsius}, using Factsage 6.3 with FT oxid and SGPS databases.}
\begin{small}\begin{tabular}{c c c}
    \hline
     Equation No.&Equilibrium equations&$\Delta$G$_{653.15K}^0$ (kJ/mol) \\
     \hline
     (4)& $4\mathrm{AgCl(s)} + \mathrm{Ti(s)} + 2\mathrm{H}_2\mathrm{O(g)}\rightarrow 4\mathrm{Ag(s)} + \mathrm{TiO}_2\mathrm{(s)} + 4\mathrm{HCl(g)}$&-430.29 \\
     (6)&$\mathrm{TiCl_4}(g)+ \mathrm{H_2O}(g) \rightarrow \mathrm{TiO_2}(s) + 4\mathrm{HCl}(g)$&-111.53\\
     (7)& $\mathrm{TiCl_4}(g)+ \mathrm{O_2}(g) \rightarrow \mathrm{TiO_2}(s) + 2\mathrm{Cl_2}(g)$&-140.57  \\
     (8)&$\mathrm{Ti(s)} + 2\mathrm{Cl_2}(g) \rightarrow \mathrm{TiCl_4}(g)$ &-683.99 \\
     (11)& $\mathrm{Sn}(s) + 2\mathrm{Cl_2}(g) \rightarrow \mathrm{SnCl_4 }(g)$ &-391.68 \\
     (12)& $\mathrm{Sn}(s) + \mathrm{Cl_2}(g) \rightarrow \mathrm{SnCl_2 }(l)$ &-246.51\\
     (13)&$\mathrm{SnCl_4}(g)+ 2\mathrm{H_2O}(g) \rightarrow \mathrm{SnO_2}(s) + 4\mathrm{HCl}(g)$&-20.23  \\
     (14)&$\mathrm{SnCl_4}(g)+ \mathrm{O_2}(g) \rightarrow \mathrm{SnO_2}(s) + 2\mathrm{Cl_2}(g)$&-49.26  \\
     (16)& $\mathrm{Al}(s) + \frac{3}{2}\mathrm{Cl_2}(g) \rightarrow \mathrm{AlCl_3 }(g)$ &-558.201\\
     (17)&$\mathrm{AlCl_3}(g)+ \frac{3}{2}\mathrm{H_2O}(g) \rightarrow \mathrm{Al_2O_3}(s) + 3\mathrm{HCl}(g)$&-155.4 \\
     (18)&$\mathrm{AlCl_3}(g)+ \frac{3}{2}\mathrm{O_2}(g) \rightarrow \mathrm{Al_2O_3}(s) + \frac{3}{2}\mathrm{Cl_2}(g)$&-354.36 \\  
     \hline
\label{table4}
\end{tabular}\end{small}
\end{table*}

\subsection{Mechanisms} 
\subsubsection{Crack initiation}
A layer of self-healing and adhesive oxide is usually formed on the top of Ti due to its high affinity with O, mainly forming rutile TiO$_2$, which can give rise to good corrosion resistance in most circumstances \cite{lutjering2007titanium,hanaor2011review}. When the passive oxide film is broken down, this exposes the underlying alloy to the salt, which always competes with a repassivation process if oxygen is accessible.  In the case of NaCl-associated HSSCC, it has been proposed~\cite{Petersen1971,joseph2018mechanisms,pustode2014stress} that TiO$_2$ can be consumed through the reaction with NaCl and moisture at elevated temperature, forming sodium titanates and gaseous HCl. A localized HCl-rich environment can retard oxide repassivation \cite{soltis2015passivity}, especially at lower oxygen concentrations underneath the salt deposits, which can enable attack of the base alloy and thence to crack initiation under applied stress. However, the reaction of AgCl with TiO$_2$ either in the presence of H$_2$O or in dry air are not energetically favoured according to equilibrium thermodynamic calculations using the Factsage database. 

Supplementary corrosion experiments performed without applied stress result in enhanced oxidation across the sample surface with the formation of oxide blisters (see Appendix), but with no cracks growing into the underlying metal surface. It is well understood that a compressive stress could be generated in the convex side of growing oxide film on a metal substrate due to the Pilling-Bedworth volume differences, although the net stress after cool-down in the oxide could be either tensile or compressive, depending on the oxide thickness~\cite{stringer1970stress}. This can give rise to mechanical damage of the oxide film, such as flaking, blisters or cracks. Therefore it is inferred that stress corrosion cracking in the present case occurs by the formation of a thick oxide caused by corrosion that is then ruptured mechanically. Such considerations may give rise to an additional tensile stress (over and above the applied stress) at the metal/oxide interface, which may facilitate crack initiation underneath the blister.

Once the protective oxide film is disrupted, it is proposed that AgCl can react with the underlying Ti in the presence of moisture and generate metallic Ag, TiO$_2$ and gaseous HCl at \SI{380}{\celsius}, reaction (4). The standard formation enthalpy for proposed reactions is present in Table 4. 
\begin{equation}
\centering 4\mathrm{AgCl(s)} + \mathrm{Ti(s)} + 2\mathrm{H}_2\mathrm{O(g)}\rightarrow 4\mathrm{Ag(s)} + \mathrm{TiO}_2\mathrm{(s)} + 4\mathrm{HCl(g)}
\end{equation}

It is commonly suggested that the source of moisture may be from water vapour in the atmosphere, adsorption on TiO$_2$ \cite{doi:10.5006/0010-9312-33.7.252} or hydrous salt particles \cite{rideout1966basic}. Thermodynamically, forming silver titanates or oxides is not favoured. SEM-EDX point analysis (Table~\ref{table1}) supports this view; the Ag-rich layers observed consisted mostly of metallic phase with very limited concentrations of oxygen.  Moreover, the porosity in both the oxide blister and in the crack mouth both indicate that gas release has occurred, such as of volatile metal chlorides as discussed in the following sections.


HCl then attacks the bare Ti alloy at the crack tip, producing titanium chlorides and atomic hydrogen, reaction (5). 
\begin{equation}
\centering \mathrm{Ti}(s) + 4\mathrm{HCl}(g) \rightarrow \mathrm{TiCl_4}(g) + 4\mathrm{H}
\end{equation}
The stability of the TiCl$_2$ and TiCl$_4$ phases has been examined\cite{chapman2015environmentally}; TiCl$_2$ is a solid phase at \SI{380}{\celsius}, while TiCl$_4$ is quite volatile, with a lower formation energy.  Volatile TiCl$_4$ can be hydrolysed or oxidised and release gaseous HCl or Cl$_2$, reactions (6) and (7). 
\begin{equation}
\centering \mathrm{TiCl_4}(g)+ 2\mathrm{H_2O}(g) \rightarrow \mathrm{TiO_2}(s) + 4\mathrm{HCl}(g)
\end{equation}
\begin{equation}
\centering \mathrm{TiCl_4}(g)+ \mathrm{O_2}(g) \rightarrow \mathrm{TiO_2}(s) + 2\mathrm{Cl_2}(g)
\end{equation}
Chloride ions (in HCl) are aggressive and able to re-attack the base alloy. 

Excess corrosion products are found to deposit into a shallow pit-like region, as observed in Figure 5(a). The brittle oxide is found to crack under the tensile stress, while the ductile Ag-rich layers above it remain continuous, which may slow down the evaporation of HCl or Cl$_2$. Since the ratio between penetration and the lateral corrosion depth is less than 1, pit formation is not considered in this case \cite{2017Sc:m}. Crack initiation is thought to be due to the slip-dissolution process under tensile stress.  It is proposed \cite{blackburn1969metallurgical} that slip promoted by hydrogen adsorption into the metal can expose material on the slip line to the corrosive environment, resulting in localized anodic dissolution. A corrosion tunnel can then be formed, which can then act as the crack initiation site. \textcolor{black}{The fractographic appearance is one of microstructure-sensitive cracking, implying a role for the morphological and/or crystallographic orientations of the underlying $\alpha$ and $\beta$ phases.} 

\subsubsection{Crack propagation}
Once a short crack initiates, the cracking or embrittlement mechanism greatly depend on the crack tip chemistry. As discussed in the preceding section,  equations (5)-(7) continually react and expose Ti at the crack tip, regenerate gaseous HCl or Cl$_2$ and supply H into the metal. The competition between the hydrolysis reaction (6) and oxidation reaction (7) is suggested to be dominated by pressure according to Chevrot's study on pressure effects on the lives of IMI834 specimens undergoing NaCl HSSCC \cite{chevrot1994pressure}. He found that the specimen life was remarkably increased at higher pressure; it was speculated that high pressure suppressed the hydrolysis of metal chlorides and furthermore restrained hydrogen charging. So for laboratory fatigue testing under relatively low pressures, it has been suggested that pyrohydrolysis reactions are favoured over oxidation, giving rise to HCl formation \cite{chapman2015environmentally,saunders2016understanding}.  Nevertheless, reaction (8) with chlorine gas is in principle thermodynamically favoured, especially at higher temperatures \cite{ciszak2016nacl,CISZAK2020108611};
\begin{equation}
\centering \mathrm{Ti(s)} + 2\mathrm{Cl_2}(g) \rightarrow \mathrm{TiCl_4}(g)
\end{equation}
This reaction will provide an intermediate corrosion product for reaction (6) and (7), furthermore driving the process of HSSCC. 

Although repassivation can occur within the crack rapidly if a threshold oxygen partial pressure is reached, the newly formed oxide film can be ruptured again at the crack tip under applied stress, exposing the bare metal through the looping reactions mentioned above \cite{chapman2015environmentally}.  Localized anodic dissolution then assists crack advance. Meanwhile, the constant supply of HCl leads to hydrogen charging at the crack tip.  Hydrogen is suggested to concentrate at the crack tip owing to the stress field and/or the existence of dislocation traps in the crack tip plastic zone \cite{garfinkle1973electrochemical}, retarding the diffusion of H into the bulk. \textcolor{black}{However, the amount must still be lower than the solubility for H in the $\beta$ at room temperature, as hydrides are not observed (see, e.g.~\cite{chang2018characterizing}).} Hydrogen embrittlement is widely proposed as the cracking mechanism in chloride-induced HSSCC \cite{rideout1966basic,Petersen1971,joseph2018mechanisms,cao2017mechanism,doi:10.5006/0010-9312-33.7.252}. \textcolor{black}{As mentioned before, the rationale of the HELP theory is that solute hydrogen atoms can interact with the elastic stress field around dislocations and increase dislocation mobility through shielding dislocation interactions \cite{birnbaum1994hydrogen, robertson2001effect}. In contrast, the mechanism of AIDE suggests that the absorption of hydrogen atoms ahead of crack tip can weaken the interatomic bonds for a few atomic layers and then facilitate the formation of dislocation cores and surface steps, leading to dislocation emission ahead of the crack tip and thus, embrittlement \cite{lynch1988environmentally,chapman2016dislocation,barrera2018understanding}. These mechanisms of non-hydride formation are likely to dominant at this temperature of \SI{380}{\celsius} rather than precipitating hydrides owing to the increased solubility of hydrogen. It is believed that the $\beta$ phase can accommodate more hydrogen than the $\alpha$ phase according to the Ti-H phase diagram and previous study on measurements of solute hydrogen in Ti-6246 by atom probe tomography (APT) \cite{san1987h,chang2018characterizing}.}
Brittle titanium hydrides were also detected by XRD and TEM in a previous study on HSSCC of Ti-6246 induced by NaCl \cite{joseph2018mechanisms}, which suggested precipitation of titanium hydrides occurs during cooling owing to the decreasing hydrogen solubility in the bulk. 

The brittle crack appearance in the $\alpha$ phase was observed in this work, which is similar to the previous findings on NaCl HSSCC of Ti-6246 \cite{joseph2018mechanisms,chapman2015environmentally}. It has been proposed that both HELP and AIDE could lead to a transgranular fracture appearance \cite{shih1988hydrogen,cao2017mechanism,chapman2016dislocation}. However, dense dislocation structures were normally found a few \si{\micro}m below the $\alpha$ facets in previous TEM studies \cite{cao2017mechanism,chapman2016dislocation}, in comparison with the observation of high dislocation density near the fracture surface ($<$ \SI{1}{\micro\meter}) in this case.  Thus the higher dislocation emission ahead of crack tip is expected during crack propagation on AgCl HSSCC. The underlying reasons might relate to the flow or migration of corrosion products within the crack, and rapid dissolution rate at the crack tip, which will contribute to facilitate dislocation emission together with the effects of hydrogen. \textcolor{black}{Conversely, Liu et al. \cite{liu2009local} also suggested that enhanced emission and mobility of dislocations could affect the local anodic dissolution by generating an additional negative potential based on the local additional potential (LAP) model.}

In this work, we have observed the presence of metallic Ag along the crack walls, generated by the corrosion reactions. Stoltz and Stulen \cite{stoltz1978solid} observed that Ag can cause solid metal-induced embrittlement on Ti-6Al-6V-2Sn above \SI{232}{\celsius}. The mechanism of solid metal embrittlement (SME) is not yet fully understood, but it has been hypothesised that the embrittler could be adsorbed at crack tips and weaken the interatomic bonds of the base metal, which then promote brittle fracture through enhancing dislocation emission or decohesion ahead of crack tips \cite{gordon1982mechanisms, kamdar1984solid,lynch1992metal}. Gordon \cite{gordon1978metal} pointed out that surface self-diffusion was likely to be the transport mechanism of embrittlers on solid metals.  The crack propagation rate would then be limited by the rate of diffusion of Ag to crack tips. Since the repassivation on crack walls can happen rapidly, Ag is expected to diffuse along the top surface of the oxide film and form islands according to STEM-EDX mapping results in Figure 7(b). The rate of diffusion of Ag on smooth and uniform TiO$_2$ surfaces has been found to be $10^{-18}\textendash10^{-21} $m$^2$/s at \SI{400}{\celsius}, based on measurements in previous work~\cite{kulczyk2014investigation,zhang2016fast}. However, vapor transport of HCl and Cl$_2$ to the crack tip is thought to be faster than the solid-state diffusion of Ag, and the presence of Cl has be found to enhance SCC susceptibility. In the present case, we speculate that crack extension precedes the surface diffusion of Ag along the oxide films formed at crack walls, based on a simplified hypothesis that there is no formation of any liquid phase mixtures containing Ag. However, intimate contact between Ag and the underlying metal could take place locally during crack incubation under at an oxygen partial pressure below some threshold value, enabling adsorption-induced solid metal embrittlement. Metal coupling interactions can be complex and remain unknown, possibly involving potential-driven electron and ion transport. Lynch \cite{lynch1992metal} also suggested that SME could produce small cracks followed by extensive cracking assisted by other modes, such as hydrogen embrittlement, stress corrosion cracking or fatigue. Consequently, our observations cannot definitively exclude a role for solid metal embrittlement.

\subsubsection{Activities of alloying elements}
In present study,  SnO$_2$ and Al$_2$O$_3$ were detected in the oxide mixture deposited above the crack mouth together with TiO$_2$, Figure 7(a). A very small amount of Cl ($<$ 1 at\%) was also found in this deposit, implying that the reactive alloying elements Sn and Al might preferentially form volatile chlorides during the reactions and then be oxidized, Reactions (9)-(18). \ce{AlCl3}, that is quite volatile at the testing temperature, is the main compound for aluminium chlorides and can react with \ce{H2O} and \ce{O2} forming \ce{Al2O3}. Sn can form two chlorides: \ce{SnCl2} and \ce{SnCl4}; \ce{SnCl4} is gaseous while \ce{SnCl2} is molten at \SI{380}{\celsius}. It was found by Rideout et al. \cite{rideout1970hot} that increasing the Al content (as an $\alpha$ stabilizer) enhances the susceptibility of Ti alloys to SCC. Beck \cite{beck1968stress} and Blackburn \cite{blackburn1969metallurgical} observed that adding $>$ 6 at\%Al in CP Ti could change the dislocation arrangements by promoting planar slip in the $\alpha$ phase, potentially owing to formation of ordered \ce{Ti3Al} $\alpha_2$. Since most titanium alloys contain these quantities of Al (or more),  this may provide a rationale for the vulnerability of Ti alloys to cracking with a brittle appearance when subjected to SCC attack. 

\begin{align}
    \ce{Sn(s) + 4HCl(g) &-> SnCl4(g) + 4H} \\
    \ce{Sn(s) + 2HCl(g) &-> SnCl2(l) + 2H} \\
    \ce{Sn(s) + 2Cl2(g) &-> SnCl4(g)} \\
    \ce{Sn(s) + Cl2(g) &-> SnCl2(l)} \\
    \ce{SnCl4(g) + 2H2O(g) &-> SnO2(s) + 4HCl(g)} \\
    \ce{SnCl4(g) + O2(g) &-> SnO2(s) + 2Cl2(g)} \\
    \ce{Al(s) + 3HCl(g) &-> AlCl3(g) + 3H} \\
    \ce{2Al(s) + 3Cl2(g) &-> 2AlCl3(g)}   \\
    \ce{2AlCl3(g) + 3H2O(g) &-> Al2O3(s) + 6HCl(g)} \\
    \ce{2AlCl3(g) + 3O2(g) &-> 2Al2O3(s) + 3Cl2(g)} 
\end{align}

STEM-EDX analysis of the Ag-rich particles, Figure 6(a) and Table 2, found substantial oxygen contents and small amounts of Cl alongside elevated amounts of Al, Sn and Mo. It might be speculated that various volatile chlorides including TiCl$_4$ could evaporate and condense on the Ag-rich particles, generating oxides and/or oxychlorides. The increasing content of Cl with increasing crack depth is considered as an additional evidence of a role for volatile chlorides. Substantial Sn contents were only found near the crack mouth, which might be attributed to the high volatility of SnCl$_4$. Although corrosion products containing zirconium  were not identified in this work, formation of zirconium oxychlorides and oxides were both observed in previous studies on NaCl-induced corrosion \cite{joseph2018mechanisms,CISZAK2020108611}. In terms of the activity of Zr and Mo, Ciszak et al. pointed out that both Zr and Mo can react with Cl$_2$ and form corresponding chlorides/oxides at \SI{560}{\celsius} based on theoretical thermodynamic modelling.  In any case, there does seem to be substantial evidence for alloying effects on the SCC susceptibility of Ti alloys, due to a combination of reactivity and volatility of the different chlorides.

\section{Conclusions}
AgCl-induced stress corrosion cracking of $\alpha+\beta$ alloy Ti-6246 used for compressor discs has been examined at \SI{380}{\celsius} and \SI{500}{\mega\pascal}. The following conclusions are addressed.

1. Metallic Ag phase was found on the metal surface after corrosion through SEM-EDX, and XRD ,which is consistent with thermodynamic analysis. This Ag contained dissolved Ti, Al and O. Ag was then observed to migrate along the crack walls, forming discrete particles containing a mixture of alloy oxides or chlorides. 

2. The protective TiO$_2$ film is suggested to be ruptured because of mechanical damage, allowing contact between AgCl and the underlying alloy. Crack initiation can be explained by a slip-dissolution model, as previously proposed. However, the effect of solid metal embrittlement on crack initiation could not be excluded by the present work.

3. TiO$_2$ in the form of anatase, SnO$_2$ and Al$_2$O$_3$ were identified by TEM study. Volatile chlorides, such as TiCl$_4$, SnCl$_4$ and AlCl$_3$, are expected to generate as intermediate corrosion products and then form oxides deposited above the crack mouth.

4. The fracture surface has a \textcolor{black}{transgranular nature with a brittle appearance on the primary} $\alpha$ laths.  The dislocation appearance is distinct from previous studies on NaCl HSSCC, consisting of a larger amount of long and straight dislocation segments well below the fracture surface. This implies an additional chemical effect on the plasticity mechanisms at the crack tip along with the effects of hydrogen embrittlement.

\section*{Acknowledgements}
The authors gratefully acknowldge the provision of materials and support of Rolls-Royce plc. DD and SJ were also supported by EPSRC (EP/K034332/1). Useful conversations are also acknowledged with John Nicholls (Cranfield University) and with Richard Chater at Imperial College. 

\section*{Data Availability}
The raw data associated with this study (electron micrographs, EDX data etc) can be obtained from the authors upon request.

\bibliographystyle{model1-num-names}
\bibliography{Reference.bib}



\end{document}